\documentclass{biophys-new}
\usepackage[utf8]{inputenc}
\usepackage{graphicx}
\usepackage[colorlinks,allcolors=cyan!70!black]{hyperref}
\usepackage[printwatermark]{xwatermark}
\newwatermark[pagex={2,4,6,8,10,12,14},color=gray!20,angle=45,scale=2.5,xpos=0,ypos=0]{ACCEPTED \\ MANUSCRIPT}

\title{A novel \textit{in vitro} device to deliver induced electromagnetic fields to cell and tissue cultures}
\runningtitle{Induced EMF delivery to cells} 

\author[1,2*]{Rea Ravin}
\author[2*]{Teddy X. Cai}
\author[3]{ Randall H. Pursley} 
\author[3]{Marcial Garmendia-Cedillos} 
\author[3]{Tom Pohida} 
\author[3]{Raisa Z. Freidlin}
\author[4]{Herui Wang}
\author[4]{Zhengping Zhuang}
\author[4]{Amber J. Giles}
\author[2,5]{Nathan H. Williamson}
\author[4]{Mark R. Gilbert}
\author[2$\dag$]{Peter J. Basser}
\runningauthor{Ravin, R. \& Cai, T. X.} 

\affil[*]{Indicates equal contribution}
\affil[1]{Celoptics, Inc., Rockville, MD, USA}
\affil[2]{Section on Quantitative Imaging and Tissue Sciences \textit{Eunice Kennedy Shriver} National Institutes of Child Health and Human Development, National Institutes of Health, Bethesda, MD, USA}
\affil[3]{The Signal Processing and Instrumentation Section, Center for Information Technology, National Institutes of Health, Bethesda, MD, USA}
\affil[4]{Neuro-Oncology Branch, Center for Cancer Research, National Cancer Institute, National Institutes of Health, Bethesda, MD, USA}
\affil[5]{National Institute of General Medical Sciences, National Institutes of Health, Bethesda, MD, USA}

\corrauthor[$\dag$]{peter.basser@nih.gov}

\papertype{Article}

\begin{document}

\begin{frontmatter}

\begin{abstract}
We have developed a novel \textit{in vitro} instrument that can deliver intermediate frequency (100 -- 400 kHz), moderate intensity (up to and exceeding 6.5 V/cm pk-pk) electric fields (EFs) to cell and tissue cultures generated using induced electromagnetic fields (EMFs) in an air-core solenoid coil. A major application of these EFs is as an emerging cancer treatment modality. \textit{In vitro} studies by Novocure Ltd. reported that intermediate frequency (100 -- 300 kHz), low amplitude (1 -- 3 V/cm) EFs, which they called ``Tumor Treating Fields (TTFields)'', had an anti-mitotic effect on glioblastoma multiforme (GBM) cells. The effect was found to increase with increasing EF amplitude. Despite continued theoretical, preclinical, and clinical study, the mechanism of action remains incompletely understood. All previous \textit{in vitro} studies of ``TTFields'' have used attached, capacitively coupled electrodes to deliver alternating EFs to cell and tissue cultures. This contacting delivery method suffers from a poorly characterized EF profile and conductive heating that limits the duration and amplitude of the applied EFs. In contrast, our device delivers EFs with a well-characterized radial profile in a non-contacting manner, eliminating conductive heating and enabling thermally regulated EF delivery. To test and demonstrate our system, we generated continuous, 200 kHz EMF with an EF amplitude profile spanning 0 -- 6.5 V/cm pk-pk and applied them to exemplar human thyroid cell cultures for 72 hours. We observed moderate reduction in cell density ($<$ 10\%) at low EF amplitudes ($<$ 4 V/cm) and a greater reduction in cell density of up to 25\% at higher amplitudes (4 – 6.5 V/cm). Our device can be readily extended to other EF frequency and amplitude regimes. Future studies with this device should contribute to the ongoing debate about the efficacy and mechanism(s) of action of ``TTFields'' by better isolating the effects of EFs and providing access to previously inaccessible EF regimes.

\end{abstract}

\begin{sigstatement}
Previous work reported that intermediate frequency (100 -- 300 kHz), low amplitude (1 -- 3 V/cm) electric fields (EFs) have an anti-mitotic effect capable of inhibiting the growth rate of cancers cell \textit{in vitro} and in the clinic, although the mechanism of action remains unclear. Existing EF delivery systems use capacitively coupled electrodes placed in direct contact with the specimen holder. This direct contact generates unwanted heat and limits the amplitude and duration of EF stimulation. We have developed a novel \textit{in vitro} system capable of continuous thermal regulation and delivery of well-characterized (200 kHz, 0 -- 6.5 V/cm pk-pk), electromagnetic fields (EMFs) to cell cultures, paving the way for improved mechanistic biophysical studies.  
\end{sigstatement}
\end{frontmatter}

\section*{Introduction}

There is an extensive body of literature describing the effects of low (DC-kHz) and high (above 1 MHz) frequency electric fields (EFs) on living cells \cite{Cifra2011}. Very low frequency EFs ($<$ 2 kHz) are used to stimulate excitable cells while high frequency EFs are used to heat cells and tissues. Intermediate frequency EFs (kHz-MHz) were previously considered to have no biological effects. However, Kirson, et al. \cite{Kirson2004} reported that EFs at frequencies of $100 - 300$ kHz and amplitudes between $1 - 3$ V/cm had an anti-mitotic effect, inhibiting the proliferation rate of human and rodent tumor cell lines while having no effect on non-cancerous cells. The inhibitory effect was reported to be frequency-dependent with a peak effect at 200 kHz for glioblastoma multiforme (GBM) cell lines \cite{Kirson2004}. The inhibitory effect was also found to increase with EF amplitude \cite{Kirson2007}. At higher EF amplitudes ($\ge$ 2 V/cm), significant ($>$ 50\%) inhibition relative to temperature-matched controls was reported \cite{Kirson2004, Kirson2007}. Kirson et al. labelled these EFs as ``Tumor Treating Fields (TTFields)''.

GBM is the most common and lethal adult brain tumor. Prognosis remains poor. Despite a combined treatment plan of maximal safe aggressive surgery, chemotherapy with temozolomide, and radiation, patients’ median survival is only 14 -- 16 months from the time of diagnosis; 1 in 4 patients survive more than 2 years and fewer than 1 in 20 survive more than 5 years \cite{Batash2017, Gilbert2014}. One significant feature of GBM is the highly invasive character of the cancer cells. This makes complete surgical resection impossible, leading to a recurrence rate of almost 100\% \cite{Gilbert2014, Stupp2005}. Based on preclinical results, ``TTFields'' entered clinical trials \cite{Stupp2012, Stupp2017} and were approved by the FDA for newly diagnosed GBM cases in 2015, representing the first new GBM treatment modality in decades. ``TTFields'' have also been approved for treatment of pleural mesothelioma as of 2019 \cite{Ceresoli2019}. 

Despite ongoing, theoretical, preclinical, and clinical research, the mechanism(s) of action of ``TTFields'' is not well understood. Abnormal cell mitotic morphology including blebbing followed by apoptosis \cite{Kirson2004, Kirson2007, Gera2015} was observed when these EFs were applied to proliferating cells, suggesting a biophysical anti-mitotic mechanism of action. Computational models predicted that the penetration of EFs into cells is indeed frequency-dependent, as suggested by findings by Kirson et al. \cite{Kirson2004}. At ultra-low frequency, EFs are largely screened and do not penetrate the cell membrane, whereas in the low to intermediate (100 -- 500 kHz) frequency range, EFs are predicted to penetrate cells and could have a biophysical effect \cite{Tuszynski2016, Wenger2015a, Wenger2015b, Wenger2018}. 

In initial preclinical studies, Kirson et al. \cite{Kirson2004, Kirson2007} proposed two specific mechanisms by which penetrative intermediate frequency EFs could cause mitotic arrest and apoptosis. The first mechanism is an interaction of these EFs with the electric dipoles of tubulin dimers which results in changes in tubulin dimer rotational dynamics, disrupting the formation and function of the mitotic spindle \cite{Davies2013, Kirson2004, Bomzon2016}. The exact effects of EFs on microtubule polymerization were predicted to depend on the length of the microtubule filament as well as the frequency and amplitude of the applied EF. Effects on septin dimers were also proposed \cite{Gera2015}. However, more recent computational modeling suggests that EFs in the range of 1 -- 3 V/cm are orders of magnitude too small to affect the rotational dynamics of tubulin and septin dimers \cite{Tuszynski2016, Wenger2015b, Wenger2018, Berkelmann2019, Li2020}. 

The second mechanism of action proposed by Kirson et al. \cite{Kirson2004, Kirson2007} was dielectrophoretic (DEP) forces resulting in electrokinetic effects on polarizable bio-macromolecules (i.e., those capable of producing a counter-ion cloud). During cytokinesis, a cleavage furrow appears between the dividing daughter cells. Computational models \cite{Wenger2015b} predict that in and around these furrows, non-uniform EFs (i.e., EF gradients) produce DEP forces on charged, polarizable macromolecules and organelles \cite{Wenger2015b, Berkelmann2019}. These DEP forces could potentially cause molecules to migrate toward or away from the furrow, impeding cell cleavage \cite{Tuszynski2016, Wenger2015b, Berkelmann2019}. Modeling studies incorporating Stokes drag \cite{Li2020}, however, suggests that DEP forces are too small to cause substantial displacement over the course of telophase. Both of these proposed mechanisms also depend on the relative orientation of the dividing cells and the applied EF. Parallel alignment results in a maximal effect \cite{Wenger2015b, Wenger2018, Kirson2007}. 

Recent preclinical studies support additional potential mechanisms and downstream effects, including but not limited to mitotic checkpoint inhibition \cite{Kessler2018}, tubulin mediated conductance effects \cite{Santelices2017}, permeability changes \cite{Chang2018}, DNA damage \cite{Giladi2017, Jo2018}, migratory inhibition \cite{Silginer2017}, and autophagy \cite{Kim2019, Silginer2017}. More recently, Neuhaus et. al \cite{Neuhaus2019} suggested voltage-gated calcium channels and Li et al. \cite{Li2020} suggested tumor cell membrane potential as mediating factors for the anti-proliferative effect of intermediate frequency EFs. While several biophysical mechanisms are plausible, none have been conclusively demonstrated.

To advance their preclinical research on the use and mechanism(s) of action of ``TTFields'', Novocure Ltd. developed an \textit{in vitro} test system (Inovitro\texttrademark) designed to deliver EFs to cell cultures using two pairs of capacitively coupled electrodes insulated by a high dielectric constant ceramic \cite{Kirson2007}, placed orthogonal to each other inside a cell culture dish \cite{Giladi2015, Kirson2007}. The electrode pairs are connected to a high voltage sinusoidal waveform generator. The stimulated electrode pair is switched periodically to maximize the orientational efficiency of EF application \cite{Kirson2007}. While contact between the culture dish and electrodes creates a direct means for EFs to be delivered to cell or tissue cultures, it also results in heat conduction from the electrodes to the dish and subsequently from the dish to the culture media, making continuous temperature control problematic. Joule or Ohmic heating from the EFs themselves also significantly contributes to heating of the conductive cell media. 

In order to maintain a temperature of $37{}^{\circ}$C in the culture dish, EF application with the Inovitro\texttrademark\ device is performed in an 18 ${}^{\circ}$C incubator \cite{Porat2017}. Electric current limits and on-off operation are also imposed as necessary to maintain temperature \cite{Porat2017, Kirson2007}. A consequence of keeping the apparatus in this refrigerated environment is a lower surrounding water vapor pressure, which accelerates cell media evaporation. Assuming 95\% relative humidity in the incubator, a loss of about 0.2 mL of media per day can be expected from a 35 mm diameter dish (see Supp. Sec. 1). As a result, media replacement is required every 24 hours according to procedures for operating the Inovitro\texttrademark device \cite{Kirson2004, Porat2017}. This rapid media loss and replacement may have inadvertent confounding effects, such as causing time-varying hyperosmotic bath conditions. Hyperosmotic conditions can result in nuclear lamina buckling \cite{Finan2010}, cell cycle disruption \cite{Radmaneshfar2013}, or protein aggregation \cite{Fragniere2019}. Another issue is that the geometry of the EF delivery device does not lend itself to easily predicting the electric field or current density produced within the tissue or cell culture \cite{Wenger2015a, Wenger2018}.    

To overcome some of these limitations associated with delivering EFs via contacting electrodes, here we describe a novel device that uses electromagnetic induction to deliver EFs/electromagnetic fields (EMFs) \textit{in vitro}, like those generated by Transcranial Magnetic Simulation (TMS) devices \cite{Bickford1965} but in a higher frequency range. More specifically, we propose the use of an air-core double solenoid coil in an LC resonant circuit to induce intermediate frequency EMF in cell or tissue cultures placed within the coil. This non-contacting method of EMF delivery thermally isolates the current-carrying coil from the culture dishes. Temperature uniformity and osmolarity control of the culture medium is improved. Greater power delivery is made possible, providing potential access to high, previously inaccessible EF amplitudes. Furthermore, the radial, linearly varying amplitude of the induced circumferential EF within the coil allows each culture dish to be exposed to a known and well-characterized range of EF amplitudes or ``doses''. Induced EMFs represent a promising alternative EF delivery method for \textit{in vitro} experimentation and may better elucidate the biophysical mechanisms of action of ``TTFields'' than the current state-of-the-art.  

\section*{Materials and Methods}

\subsection*{200 kHz induced EMF generation}

\begin{figure} [hbt!]
    \begin{center}
    \includegraphics{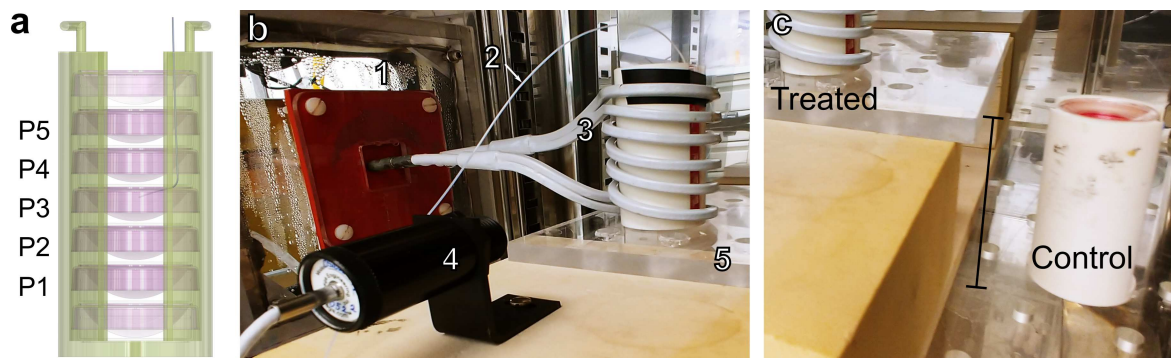}
    \caption{Experimental apparatus inside incubator. (\textbf{a}) Rendering of dish sleeve containing stack of seven 35 mm cell culture dishes. Vertical positions labelled P1 - P5. A fiber optic temperature sensor is placed into the P3 dish. (\textbf{b}) Inside view of modified incubator and experimental apparatus. (1) 22 cm $\times$ 22 cm Plexiglas window. (2) Fiber optic sensor. (3) Double-wrapped copper coil, 3 turns, ID $ = 5$ cm, height $= 8$ cm, attached to induction device. (4) Infrared detector. (5) Plexiglas shelf. (\textbf{c}) Relative position of control stack, off-axis from the treated sleeve to avoid the $1/r$ 
    EMF contribution outside the coil -- see Eq. \eqref{eqn: EMF profile}.}
    \label{fig:coil and dish stack}
    \end{center}

\end{figure}

EMF generation was accomplished using an industrial 10 kW induction heater system (DP-10-400, RDO Induction LLC, Washington NJ) based on a conventional LC resonant circuit design (with a resonant angular frequency of 250 kHz). The induction device was connected to a copper coil built to the following specifications: inner diameter (ID) of 5 cm, height of 8 cm, 3 turns, double-wrapped, with vertical orientation (Fig. \ref{fig:coil and dish stack}b, 3). The induction system and coil were jacketed and connected to a water chiller (DuraChill\textcopyright\, DCA200, PolyScience, Niles, IL) to remove heat and control coil temperature. In order to place the coil inside the 95\% air, 5\% CO${}_2$ incubator at $37{}^{\circ}$C (MCO-18M Multigas Incubator, SANYO Electric Co. Ltd., Osaka, Japan), a 22 cm x 22 cm hole (Fig. \ref{fig:coil and dish stack}b, 1) was made in the incubator to avoid Joule heating of the metal lining via induction currents passing through the copper pipes connecting the coil to the induction device. The hole was covered and sealed with Plexiglas connected to the incubator. A Plexiglas shelf was placed below the coil (Fig. \ref{fig:coil and dish stack}b, 5), replacing the metal shelf of the incubator, again to avoid inductive heating. Plastic sleeves for cell culture dishes were made in-house on a 3D printer (Fig. \ref{fig:coil and dish stack}a). Each sleeve accommodates seven 35 mm cell culture dishes in a vertical stack, so several culture dishes can be studied in parallel. The middle 5 dishes are situated fully within the coil and their positions are labelled P1 -- P5. The dish in the third position from the bottom is referred to as P3, and is shown with a fiber optic temperature sensor placed in its center (Fig. \ref{fig:coil and dish stack}a). An infrared detector records the coil temperature (Fig. \ref{fig:coil and dish stack}b, 4). EMFs were delivered continuously for 72 hours (h) in all experiments.

The expected EF profile generated within the coil by the induction device can be calculated from basic principles of electricity and magnetism. The magnetic field within an air-core current-carrying coil is readily approximated by Ampere's law,
\begin{equation}\label{eqn: B}
    B_z(t) = \mu_0 I(t) \left(\frac{N}{L}\right)
\end{equation}
where $B_z(t)$ is the axial component of the applied magnetic field, $\textbf{B}$, $\mu_0$ is the permeability of a vacuum, $I$ is the current amplitude, $N$ is the number of coil turns, and $L$ is the length of the coil. The direction of the magnetic field is normal to the direction of the current and can be determined by the right hand rule as being along the coil axis. Faraday's law in integral form relates the magnetic field to the electromotive force, $\varepsilon$,
\begin{equation}
    \varepsilon = \oint_l \textbf{E} \cdot d\textbf{l} = - \frac{d}{ dt } \iint_S \textbf{B} \cdot d\textbf{S}, 
\end{equation}
where $\textbf{E}$ is the induced EF, $\textbf{S}$ denotes some bounded surface, and $d\textbf{l}$ is an infinitesimal arc length. If the surface is chosen to be a circular cross section of the coil with radius $r$ from the center of the coil then the following solutions are obtained for $|\textbf{E}(t)|$,
\begin{equation}\label{eqn: EMF profile}
|\textbf{E}(r,t)| = |E_\theta(r,t)| =
\begin{cases}
    \left(\dfrac{r}{2}\right) \dfrac{dB_z(t)}{dt} = \mu_0 \left(\dfrac{N}{L}\right)\left(\dfrac{r}{2}\right) \left| \dfrac{dI}{dt} \right| & r < R, \\ \\
    \left(\dfrac{R^2}{2r}\right) \dfrac{dB_z(t)}{dt}   = \mu_0 \left(\dfrac{N}{L}\right)\left(\dfrac{R^2}{2r}\right) \left| \dfrac{dI}{dt} \right| & r\ge R,
\end{cases}
\end{equation}
where $R$ is the radius of the coil, and $E_\theta(r,t)$ is the azimuthal component of the electric field. By symmetry, the other two components of the electric field vanish. The direction of the EF is anti-parallel to the current flowing in the coils. If the applied current is sinusoidal, e.g., $I(t) = I_0 \cos(\omega t)$, then the induced EF will oscillate at the same frequency in the circumferential direction. By Eq. \eqref{eqn: EMF profile}, our induction system delivering current at 200 kHz results in a circumferential 200 kHz EF profile within the coil with a linearly increasing amplitude from the center of the coil. Importantly, the radial profile of the generated EFs makes each treated dish its own ``dose titration'' experiment, with little to no EFs dose or amplitude in the center of the dish and high dose at the periphery. The effects of EF amplitude can be isolated from any dish-to-dish experimental confounds by analyzing radially dependent differences \textit{within} a single treated dish.

Note, the electric field outside of the coil is expected to decay by $1/r$ according to Eq. \eqref{eqn: EMF profile}. The EF in the incubator outside the coil is therefore not 0. Because of this decaying EF, the control dish sleeve is placed in the incubator at a distance and vertical height offset (off axis from the coil) such that little to no EF should be present, as shown in Fig. \ref{fig:coil and dish stack}c. 

\subsection*{Temperature monitoring and control}

Joule heating remains a major hurdle to delivering increased EF amplitudes while maintaining controlled temperatures, even in the absence of conductive heating from contacting electrodes. Heating of an electrically conductive material by a spatially-varying EF is described by the power delivered per unit volume
\begin{equation}\label{eqn: dPdV}
\begin{gathered}
    \frac{d P}{d V} = \textbf{J}\cdot \textbf{E} = \sigma |\textbf{E}|^2 \\  \mathrm{by}\;\;\; \textbf{J} = \sigma \textbf{E} \;\;\;\; \mathrm{(microscopic\; Ohm^{\prime}s \;Law)},
\end{gathered}
\end{equation}
where $P$ is power, $V$ is volume, $\textbf{J}$ is current density, $\textbf{E}$ is the EF, and $\sigma$ is the conductivity of the material. In the case of a solenoid coil (cylindrical geometry) surrounding some uniform conductive material, the rate of heat generation, $\dot{q}$, delivered by the current-carrying coil to the material of height, $L_z$, can be calculated from the within-coil part of Eq. \eqref{eqn: EMF profile} and Eq. \eqref{eqn: dPdV} as
\begin{equation}\label{eqn: q}
   \dot{q} \equiv P_{\mathrm{total}} = \sigma\iiint_V  |\textbf{E}(r)|_{\mathrm{rms}}^2 \, dV = \sigma \int_0^{L_z} dz \int_{0}^{2\pi}d\theta \int_0^R r  |\textbf{E}(r)|_{\mathrm{rms}}^2 \,dr    =   \frac{ \pi \sigma L_z m^2 R^4}{16},
\end{equation}
where $\iiint_V dV$ is a volume integral and $m$ is the radial slope of the within-coil part of Eq. \eqref{eqn: EMF profile} using peak-to-peak amplitudes such that $|\textbf{E}(r)|_{\mathrm{rms}} = mr/(2\sqrt{2})$,
\begin{equation}\label{eqn: E slope}
    m := \frac{\mu_0}{2} \left(\frac{N}{L}\right) \left| \frac{dI}{dt}\right|_{\mathrm{pk-pk}} \; [=] \; \mathrm{V}\cdot\mathrm{m}^{-2},
\end{equation}
where rms is the root-mean-square. According to Eqs. \eqref{eqn: dPdV}, \eqref{eqn: q}, and \eqref{eqn: E slope}, Joule heat generation increases with the square of the delivered EF amplitude. Thus, delivering higher EF amplitudes while maintaining well-controlled temperatures requires careful regulation of environmental conditions. Not only does the current-carrying solenoid coil require cooling, the cell cultures themselves require negative heat flux to the surroundings to balance Joule heating. This negative heat flux is attained by cooling the coil \textit{below} the incubator temperature. A complex interplay of incubator temperature, a cooled coil temperature, and various thermal resistances (Fig. S2, Table S2) determines the steady-state temperature profile within the cell culture dishes. See Supp. Sec. 2 for a more complete discussion of heat transfer and expected temperature profiles (Figs. S3 -- S5).

A data acquisition and analysis system was used to monitor the environmental conditions within the incubator during experiments. The system can acquire data from as many as four temperature probes and two electric field sensors. The data acquisition software was developed using LabVIEW (National Instruments, Austin, TX). During operation, the system collected temperature data from Opsens multi-channel signal conditioners (TempSens TMS-G2-1--100ST-M1, Opsens, Quebec, Canada). Data from sensors was measured every second and transmitted to the system computer via an RS-232 interface. An infrared detector was used to monitor the coil temperature. Fiber optic sensors (OTG-A-10-62ST-1, Opsens, Quebec, Canada), rather than traditional thermistors, were used to measure temperature within the cell media to avoid both inductive heating effects and any parasitic inductance and capacitance potentially contributed by thermistors. Fiber optic sensors were attached to dishes (e.g., Fig. \ref{fig:coil and dish stack}b, 2) using bone wax (Lukens\texttrademark \,\#901, Surgical Specialties, Westwood, MA). 

The water chiller set-point and temperature were also monitored and controlled. The water temperature of the chiller was measured once per second and transmitted to the system computer using an RS-232 interface. The temperature set-point of the chiller was managed by the controller software and was adjusted based on the fiber optic temperature measurements from the center of the P3 control dish. When the measured temperature deviated from the desired range, an alarm was triggered and the set-point temperature of the water chiller was adjusted remotely via open-loop control. The status of the system was monitored remotely using a cloud service (Blynk Inc., New York, NY). The LabVIEW software communicated with the Blynk cloud service to transmit real-time data. A Blynk cellphone app was used to receive data from the Blynk cloud and provide the status of the ongoing experiment. The app also had the ability to transmit data through the Blynk cloud to remotely adjust the chiller set-point in real-time in order to maintain stable temperature conditions throughout 72 h experiments.

\subsection*{Osmolarity regulation}

Cooler air near the dishes can result in accelerated loss of media due to water vapor pressure gradients (see Supp. Sec. 1). As a result, osmolarity can change over the course of experiments. We opted to take two preparatory steps to mitigate osmolarity changes, as opposed to replacing cell media during the course of the experiment. First, we used 3 mL of media rather than the conventional 2 mL; increasing the total volume decreases the proportion of media lost. Second, we shaped the lids of all dishes by heating the lids with a heat gun, placing them on a sharpened ring mold, and depressing them while malleable with a steel ball. The end result was a curved or domed lid that is partially submerged in the media. This lid shape mitigated media loss primarily by decreasing the exposed liquid surface area and thus the area undergoing mass transport with the surrounding air (Fig. S1). The lid  also directs condensate back into the media volume. These changes reduced the predicted osmolarity change by over 60\% (Table S1). Note, increased media volume and decreased exposed surface area should not affect dissolved gas (O${}_2$, CO${}_2$, etc.) concentrations in media. According to Henry's law, a dissolved gas in liquid is only a function of the partial pressure of the gas above the liquid and temperature, which are both unchanged by these modifications. Osmolarity measurements with a Vapro 5520 vapor pressure osmometer (Wescor, Logan, UT) were used to experimentally validate the efficacy of these regulation measures. 

\subsection*{Human thyroid cell culture}

We used a human thyroid cell line (Nthy-ori 3-1 Sigma, Sigma-Aldrich, St. Louis, MO) as an exemplar test cell culture system. Cells were infected with lentivirus green fluorescent protein (GFP) and selected for high expression of GFP over several cycles. Cells were maintained in 75ml flasks in DMEM media (Gibco, Thermo Fisher Scientific, Waltham, MA) supplemented with 10\% FBS (Gibco), 1\% Penicillin Streptomycin (Gibco). GFP-expressing cells were passaged twice a week and cell cultures were maintained in a 95\% air, 5\% CO${}_2$ incubator at 37${}^{\circ}$C. Before each 72 h experiment, cells were detached from the culture vessel with Accutase\textcopyright\, (Sigma-Aldrich) and re-suspended in media at a density of 3.33$\times$10${}^4$ cells/ml. 3 ml of cell suspension was placed in each 35 mm culture dish (Corning, Falcon) for a plating density of 1$\times$10${}^5$ cells per dish. Dishes were previously coated with Poly-D-Lysine to enhance attachment. Dishes were incubated for 24 h to allow for attachment prior to 72 h experiments. Markings were made on the external part of the dishes -- consistent relative to manufacturer markings -- to keep the orientation of dishes consistent during treatment and imaging.

A series of cell counting calibration experiments using different initial seeding densities ranging from 3$\times$10$^4{}$ to 1.2$\times$10$^5{}$ cells per dish were also performed. Cells were again detached with Accutase\textcopyright\,, plated at different densities, incubated for 24 h, and then incubated for an additional 72 h. At the end of 72 h, the cell culture dishes were scanned using confocal imaging, described below, and all cells were subsequently lifted and counted using a Countess™ II FL Automated Cell Counter (Thermo Fisher). Final cell densities ranged from $\sim 100$ to $1200$ cells/sq. mm. 3 series of calibration experiments were performed for a total of 56 cell counts and images. Results from these calibration experiments were used to relate imaging observables to the experimental cell counts. 

\subsection*{\textit{In situ} confocal imaging and ethidium homodimer-1 and propidium iodide staining}

A unique challenge of the study design is the presence of a radial dependence in the experimental conditions. This radial dependence precluded more obvious approaches like lifting and counting all cells and informed our choice to use GFP-expressing cells. Entire dishes were imaged \textit{in situ} to retain radial information, i.e., what EF amplitude treated cells experienced during the 72 h EMF application. Stitched images were acquired and analyzed.

At the end of experiments, sleeves with dishes were removed from the coil and dishes were gently removed from the sleeves so as not detach cells. Orientation information was retained using dish markings. All dishes were incubated until imaged. The dishes were mounted to the stage of a Zeiss LSM 780 laser scanning confocal microscope system. For imaging, an EC Plan-NeoFluor 5x, 0.16 N.A. was used. All the visible parts of the 35 mm culture dish were imaged using stitching mode, stitching 484 areas of interest with 5\% overlap. The final image size was $10,726 \times 10,726$ pixels (px). For GFP, a 488 nm excitation wavelength, a beam splitter of 488/562 nm, and emission filters of 500 - 555 nm were used. Special care was taken to use non-saturating parameters. Imaging parameters were kept constant across dishes and experiments. 

In 4 experiments, additional ethidium homodimer-1 (E1169, Thermo Fisher) (EthD-1) and propidium iodide (V13241, Thermo Fisher) (PI) staining was performed following GFP imaging. For PI, at the end of EF application and 5 minutes prior to microscope imaging, 750 $\mu$L of DMEM was removed from the dishes and mixed with 0.5 $\mu$L of 1 mg/mL stock solution and added back to the dish to a final concentration of $2 \times 10^{-4}$ mg/mL. This procedure was used to avoid agitating all media in the dish and losing spatial information on floating, or incompletely adhered cells. For EthD-1, an identical procedure was performed but for a final concentration of $0.4$ $\mu$M. No additional washes were performed to avoid loss of the floating cell population. For both PI and EthD-1, a 561 nm excitation wavelength and emission filters of 578 -739 nm were used. Stitching was identical. 

\subsection*{Image processing}

\begin{figure}[hbt!]
    \begin{center}
    \includegraphics{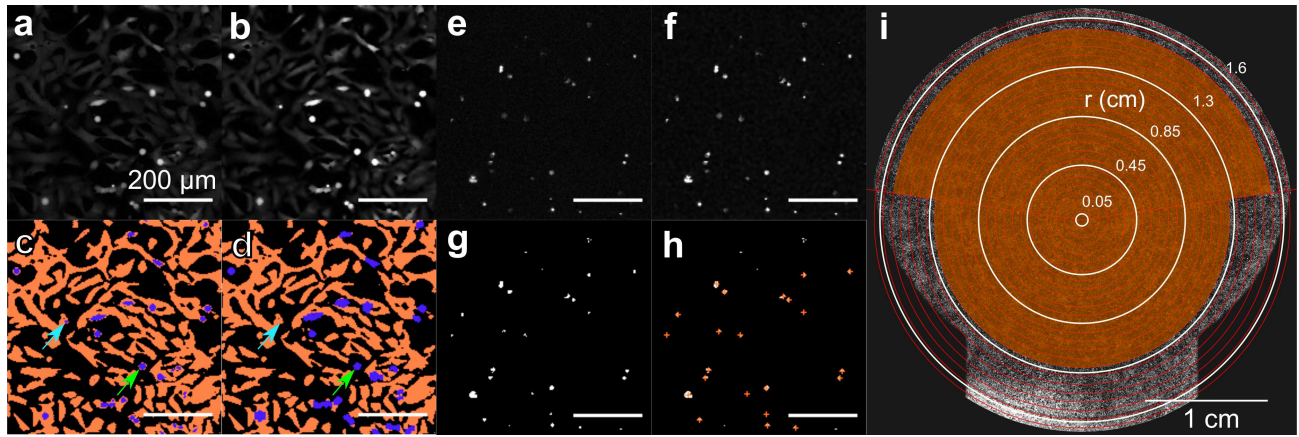}
    \caption{Overview of image processing pipeline. (\textbf{a} - \textbf{d}) Processing of GFP images. (a) Raw image. (b) Contrast enhanced image. (c) Segmented image using 3-level intensity thresholding prior to corrections. (cyan arrow) Example of ``spot'' classification error. (green arrow) Example of ``halo'' classification error. (d) Segmented image after corrections, demonstrating good agreement with qualitative segmentation of (a). (\textbf{e} - \textbf{h})  Processing of PI and EthD-1 images. (e) Raw image. (f) Contrast enhanced image. (g) Segmented image using 2-level intensity thresholding. (h) Counted particles or cells based on size thresholding. (\textbf{i}) Schematic representation of radial binning and exclusion of occluded regions on an example stitched dish image. Shaded region is included. Image has been linearly contrast enhanced ([0.01, 0.2] $\rightarrow$ [0.1, 1]) and down sampled to 20\% resolution to aid visualization.}
    \label{fig: imaging}
    \end{center}

\end{figure}

Two populations were visible in images of the GFP-expressing thyroid cells: a brighter, high fluorescence population, which might correspond to cellular debris, mitotic cells, pre-apoptotic cells, or a rounded cell morphology; and a dim, lower fluorescence population that corresponds to attached cells. The primary population of interest is the attached cell population. Therefore, one goal of image processing is to segment these bright and dim px populations for downstream analyses. This segmentation was performed in 3 steps: contrast enhancement and denoising (e.g., Fig. \ref{fig: imaging}a and b, Fig. S7), 3-level intensity thresholding (Figs. \ref{fig: imaging}c, Fig. S8), and classification error correction (Fig. \ref{fig: imaging}d, Fig. S9), outlined below. EthD-1 and PI staining images underwent a simpler procedure of contrast enhancement and denoising (Figs. \ref{fig: imaging}e and f) followed by 2-level intensity thresholding (Fig. \ref{fig: imaging}g) and size thresholding (Fig. \ref{fig: imaging}h) to identify stained dead cells. See Supp. Sec. 3 for more details and MATLAB code. After processing, images were binned radially (Fig. \ref{fig: imaging}i) to encode radial information in the analysis. Edge regions and dish regions occluded by fitments were excluded. The image processing pipeline is summarized in Fig. 2.

\subsubsection*{Contrast enhancement and denoising}

The contrast enhancement step (Figs. \ref{fig: imaging}b and f, Fig. S7, Supp. Sec. 3.1) was based on a morphological transform method \cite{Hassanpour2015, Soille2004} (Supp. Sec. 3). An image, $A$, can be contrast enhanced by computing
\begin{equation}\label{eqn: morpho contrast}
    A_{\mathrm{Enhanced}} = A + \left[A - (A\circ B)\right] - \left[(A\bullet B) - A\right] 
\end{equation}
where $B$ is a structuring element, $\circ$ denotes an opening operation, and $\bullet$ denotes a closing operation. Opening is erosion followed by dilation while closing is dilation followed by erosion. The middle term on the right-hand side of Eq. \eqref{eqn: morpho contrast} is a `top hat' of the original image, weighted to brighter regions, and the rightmost term is a `bottom hat', weighted to darker regions. Morphological contrast enhancement can therefore be conceptualized as `adding' to bright regions, and `subtracting' from dark regions \cite{Hassanpour2015}. All images underwent morphological contrast enhancement using a `disk' shaped structuring element with a 3 px radius. Operations were performed using MATLAB 2019a (MathWorks, Natick, MA) and the imtophat() and imbottomhat() functions found in the Image Processing Toolbox. After contrast enhancement, a denoising step was performed with a 3$\times$3 Wiener filter. Identical contrast enhancement and denoising was performed for all GFP, EthD-1, and PI images.

\subsubsection*{Segmentation via thresholding}

To segment the contrast enhanced GFP image, two intensity thresholds were needed: a lower threshold to separate the background and dim px population and an upper threshold to separate the bright and dim px populations (Supp. Sec. 3.2). The upper threshold (0.3727) was selected by applying Otsu's method \cite{Otsu1979} with 3 clusters to intensity histograms drawn from a central region of all control dish images from positions P2, P4, and P5 and then taking the average of the larger of the two returned intensity thresholds. The lower intensity threshold (0.0635) was selected by first filtering the contrast enhanced and denoised image region using a $3\times3$ matrix of ones, i.e., a standard deviation filter. The filtered image was then thresholded using Otsu's method with 2 clusters. The px population with intermediate standard deviation values was identified as corresponding to attached cell edges (Fig. S8a). An intensity threshold was then selected such that no more than 25\% of the identified edges were excluded by the threshold (Fig. S8c). This process was repeated for all central regions of P2, P4, and P5 control dish images. The averages of the selected lower and upper intensity thresholds (Fig. S10) were used for all images (e.g., Fig. \ref{fig: imaging}c, Fig. S9).

To segment the contrast enhanced EthD-1 and PI images, Otsu's method was used to generate a single intensity threshold. After quantizing (see Fig. \ref{fig: imaging}g), contiguous regions larger than 3 px and smaller than 50 px were counted as stained cells (Fig. \ref{fig: imaging}h). 

\subsubsection*{Classification error correction}

After applying thresholds and quantizing filters to cluster the GFP image into background, dim, and bright populations, two types of classification errors were corrected (Supp. Sec. 3.3, Fig. S9). The first was a ``halo'' effect around bright regions caused by intensity drop-off such that the edges of bright regions were incorrectly classified as dim px. The second was a ``spot'' effect in the center of dim regions caused by intensity peaks. To correct the first type of error, all bright regions exceeding 8 px in size with a surrounding px population of at least 60\% dim px were expanded by one px layer to fill the surrounding dim px (Figs. \ref{fig: imaging}c and d, green arrow), i.e., a flood fill. To correct the second type of error, all small bright regions ($<$ 8 px) that were fully enclosed by dim px were converted to dim px (Figs. \ref{fig: imaging}c and d, cyan arrow). Two passes of these corrections were performed.

\subsubsection*{Radial binning and exclusion}

Data was analyzed as a function of radius by constructing radial bins or bands. For GFP images, an arbitrary number of equal annular width bands, 25, was used to divide the total image area starting from the dish center. The choice of band number results in approximately 0.72 mm wide bands. The bottom portion of the dish is occluded by a fitment and also shows irregularities. To avoid this area, only an upper portion of the band was kept from band 17 and on. The edge of the dish was also excluded due to irregularities and observed temperature differences: bands 23 -- 25 were discarded from all analyses (Fig. \ref{fig: imaging}i). For the EthD-1 and PI stained images, a similar binning procedure with 12 bands of approximately 1.46 mm annular wide was used, with exclusion of lower regions after band 8, and a complete exclusion from band 10 onward.

\subsection*{Study design}

In each experiment, 10 dishes were simultaneously seeded. Dishes were then randomized to either the control or treated dish sleeve/stack. Pairs of dishes were assigned to the same height position (P1 -- P5). For the dishes assigned to P3, one dish is discarded and a dish with only media is used to continuously probe temperature in the treated dish sleeve. After 24 h incubation followed by 72 h of incubation and continuous 200 kHz induced EMF application to the treated dish sleeve, dishes were imaged and processed as described. Using data from cell counting calibration experiments, the imaged dim px density was related to cell density using an empirical fit, shown later. Each treated dish's radially dependent cell density was normalized to its position matched (e.g., P2 treated vs. P2 control) control dish. In 4 experiments, additional PI and EthD-1 staining, imaging, and image processing was performed to clarify the role of cell death in results. 

Each experiment generated 3 replicates of dish pairs at P2, P4, and P5. Pairs of dishes assigned to P1 were discarded from all analyses due to a substantial temperature difference, discussed later. Each dish pair was treated as a single experimental observation of the effect of 200 kHz EFs. Data from Kirson et al. \cite{Kirson2007} suggest that treatment of proliferating human cells with an EF of 200 kHz and 1 V/cm results in a decrease of roughly $20 \pm 10\%$ in cell number after 72 h compared to untreated cells. Presuming an effect of this magnitude, 18 replicates are needed to obtain a conservative statistical power of $0.95$ and a confidence level of $\alpha = 0.01$. 12 experiments were performed using our system, resulting in $N = 36$ replicates. In 2 of these experiments (6 replicates), PI staining and imaging was performed after primary imaging. In 2 other experiments, EthD-1 staining and imaging was performed for a total of $n=$ 12 out of 36 replicates assessing cell death.

\subsection*{Statistical analysis and methods}

All statistical analyses were performed in R 3.6.2. The linear fit for the experimentally measured EF amplitude was performed using the typical linear least squares (lm) approach. The bounded exponential fit for dim px density vs. cell density calibration (of the form: $[\mathrm{Dim \; Px\; Density}] \sim 1 - \exp{\{\beta_1 \times [\mathrm{Cell\; Density}] + \beta_0\}}$) was performed using the Gauss-Newton nonlinear least squares (nls) approach. Initial guesses of $\beta_1 =- 0.001$ and $\beta_2 = 0$ were provided. Bootstrapping for 95\% confidence intervals was performed using the bias-corrected and accelerated (BCa) approach provided in the R boot package (ver. 1.3-24). For bootstrapping, 1000 re-samplings were performed, following convention. Two-sample independent $t$-tests were performed using the base R t.test function assuming unequal variances, i.e., Welch's $t$-test. 

\section*{Results}

\subsection*{Delivery of 200 kHz EFs up to $\sim$ 6.5 V/cm pk-pk to cell cultures via induced EMFs}

\begin{figure}[hbt!]
    \begin{center}
    \includegraphics{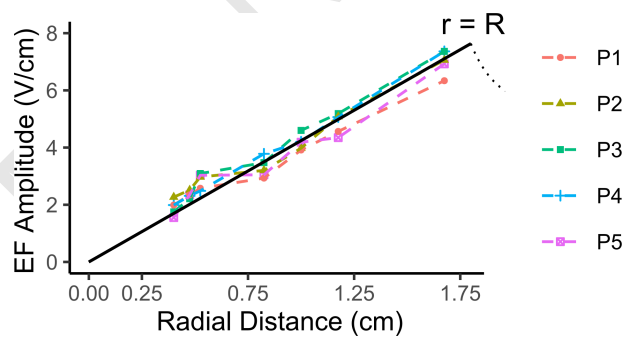}
    \caption{Measurements of EF amplitude as a function of pickup coil radius/radial distance at different dish stack positions. Black line is a zero-intercept linear least squares fit using data from all positions ([EF Amplitude] = 4.242 $\times$ [Radial Distance], Std. Error = 0.071, Adj. R${}^2$ = 0.9902, p $<$ 0.001). }
    \label{fig: EF measurement}
    \end{center}
\end{figure}

The EF amplitude and frequency delivered to the treated dish sleeve were measured using pickup coils (i.e., loops) of several different radii composed of insulated twisted wire pairs to minimize stray inductance. These pickup loops were hot glued to dishes, aligned centrally, and placed within the sleeve. The generated peak-to-peak voltage (V${}_\mathrm{pk-pk}$) was measured with a two-channel oscilloscope (Hantek6022BL, Hantek Electronic Co., Ltd., Qingdao, China) and EF amplitude was determined by $|\textbf{E}| = \mathrm{V_{pk-pk}}/(2\pi r)$ where $r$ is the loop radius. Results from these measurements are shown in Fig. \ref{fig: EF measurement}, and agree with the expected linear EF amplitude profile predicted within the coil from Eq. \eqref{eqn: EMF profile}. 200 kHz oscillation was also observed in all measurements, confirming the linearity of the system. Note, EF amplitudes throughout refer to pk-pk values. We obtained a maximal EF amplitude of $\sim$ 6 -- 7 V/cm at the periphery of the dish with a power level of 3.19 kW. The power level is a consequence of 220 VAC operation and the inductive load of the culture dishes and fixture system. We report a method of delivering EFs with a large dynamic amplitude range and well-characterized amplitude profile.

\subsection*{Temperature and osmolarity control during 72 h continuous 200 kHz EMF delivery} 

\begin{figure}[hbt!]
    \begin{center}
    \includegraphics{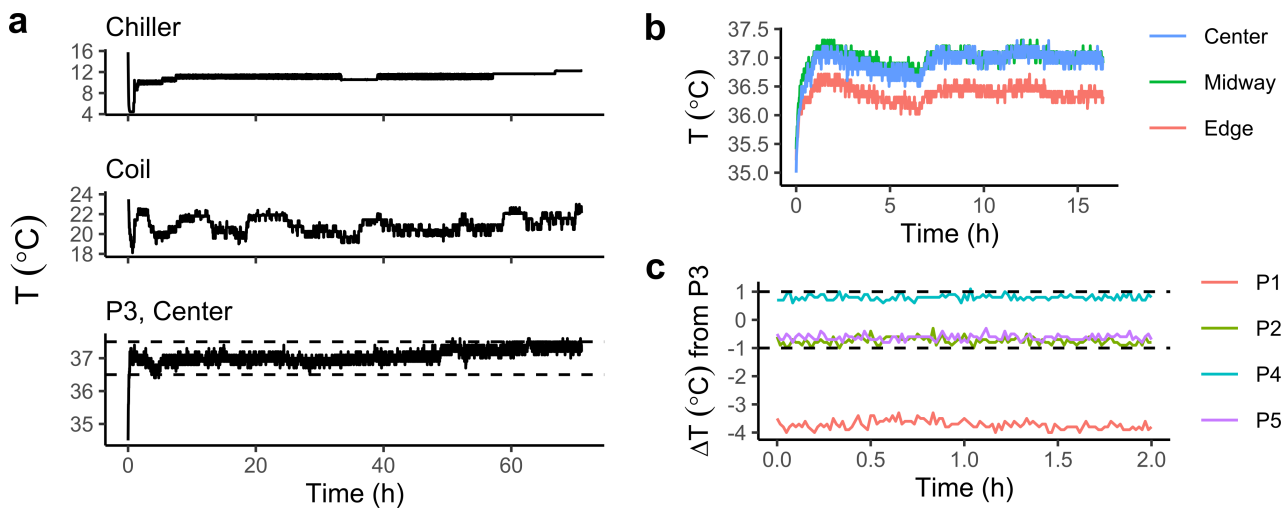}
    \caption{Quantification of temperature stability and homogeneity. (\textbf{a}) Temperature log data for the entire duration of a single 72 h experiment from (top) the chiller water loop, (middle) the infrared detector reading of the coil, and (bottom) the fiber optic sensor placed in the center of the dish at P3. Dashed lines indicate $37\pm 0.5 {}^{\circ}$C. Note the fluctuations in coil temperature and chiller set-point adjustment throughout the experiment. (\textbf{b}) Adjusted temperature data from 3 fiber optic probes placed in different positions of the dish at P3 during the first 17.5 h of an experiment. (\textbf{c}) Temperature data from fiber optic probes placed in the centers of dishes P1, P2, P4, and P5, expressed as a difference from a simultaneous reading of a probe in the center of the dish P3 with dashed lines indicating $0\pm 1 {}^{\circ}$C.}
    \label{fig: T measurement}
    \end{center}

\end{figure}

These cell cultures are typically grown at a temperature of $37{}^{\circ}$C. Extensive steps were taken to maintain this desired temperature in the treated dish sleeve as growth rate is known to be function of ambient temperature. For the present experimental setup, $m = 4.242$ V/cm${}^2$ (slope of the fit Fig. \ref{fig: EF measurement}), such that $0.1$ W $\equiv 3.6$ kWh of Joule heating is expected per dish by Eq. \eqref{eqn: q} (see Supp. Sec. 2.4, Eq. S23). To balance this heat generation, the coil is jacketed and actively cooled by an external water chiller to $\approx 22\pm 1{}^{\circ}$C, cooling the surrounding air to keep the dish at P3's temperature at a homogeneous $37\pm 0.5 {}^{\circ}$C. No inadvertent effects like Rayleigh-Bénard convection are expected using this temperature control method (Supp. Sec. 2.5, Eq. S24, Fig. S6). This method of temperature control was able to maintain the desired P3 dish temperature, without interruption and without fail, for all 72 h experiments. Exemplar temperature logs for the P3 fiber optic sensor, the infrared coil recording, and the chiller loop over the course of one experiment are shown in Fig. \ref{fig: T measurement}a.


To assess the homogeneity of temperature during a stimulation experiment, measurements from different locations within the P3 dish (Fig. \ref{fig: T measurement}b) (within-dish variability) and measurements from the center of different dish positions (Fig. \ref{fig: T measurement}c) (between-dish variability) were taken. For the measurements between different dish positions, data is presented as a difference from a simultaneous P3 dish measurement. In Fig. \ref{fig: T measurement}b, stimulation was started at time $= 0$. Temperature moves in tandem, demonstrating no time-lag in temperature adjustment. Readings show that temperature within the dish is similar, but the edge of the dish is $\approx$ 0.5${}^{\circ}$C cooler. This was likely the result of convective heat transfer with the cooler surrounding air (see Supp. Secs. 2.1 - 2.4). The edges of all dishes were excluded in downstream analyses as described in Fig. \ref{fig: imaging}i to ensure the integrity of within-plate information. In Fig. \ref{fig: T measurement}c, data is shown from segments of experiments that are underway and at steady state thermal conditions. Temperature measurements from the centers of P2, P4, and P5 remained within 1${}^{\circ}$C of measurements from the center of P3 and were stable. Measurements from the center of P1, however, were several ${}^{\circ}$C cooler and so P1 was excluded from all analyses. Overall, temperature control was satisfactory and allowed us to retain most data with confidence that temperature differences were minimal and unlikely to contribute to observed differences between experimental conditions.  

\begin{figure}[hbt!]
    \begin{center}
    \includegraphics{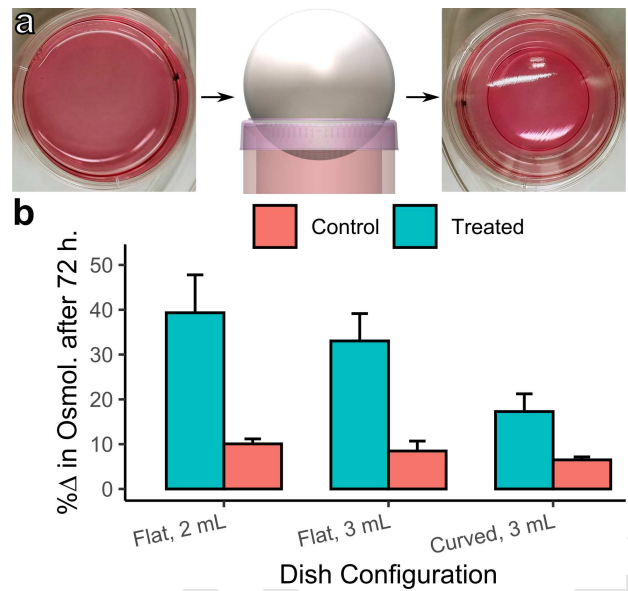}
    \caption{Quantification of osmolarity regulation improvements. (\textbf{a)} Lid shaping process with steel ball impressing a pre-heated lid placed on a sharpened ring mold. Note exposed surface area reduction in shaped lid. (\textbf{b}) Osmolarity changes after a 72 h experiment in different conditions and configurations with 3 repetitions. Error bars = $\pm 1$ SD.}
    \label{fig: osmo}
    \end{center}

\end{figure}

Osmolarity was measured after a 72 h experiment for control and treated dishes in 3 different dish configurations: (1) flat lid, 2 mL media, (2) flat lid, 3 mL media, and (3) curved lid, 3 mL media, to quantify the efficacy of the described mitigation methods (Fig. \ref{fig: osmo}, Table S1). 3 repetitions of each condition (P2, P4, P5 from one experiment) were performed. No cells were present in the media during these osmolarity experiments. Results in Fig. \ref{fig: osmo}b are expressed as a percentage change from the measured initial average osmolarity of 314 mOsm. The combined mitigation techniques reduced the osmolarity increase in the treated condition from about 39\% to 17\% and in the control condition from about 10\% to 6.5\% (Supp. Sec. 1, Eq. S10). The reduction in osmolarity increase translated to a final osmolarity of about 370 mOsm in the treated condition compared to about 335 mOsm for the control condition. The mitigation was imperfect but nonetheless a marked improvement from a final osmolarity of about 440 mOsm in the unmitigated, treated condition.

\subsection*{Estimating counted cell density using segmented px density obtained via \textit{in situ} imaging of GFP-expressing thyroid cells}

\begin{figure}[ht!]
    \begin{center}
    \includegraphics{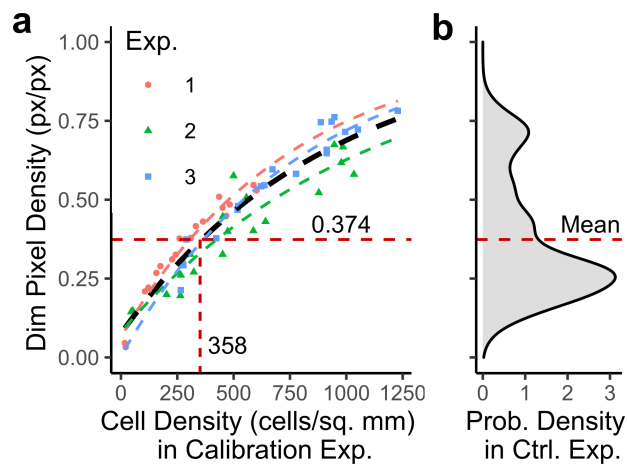}
    \caption{Translating dim px density to cell density. (\textbf{a}) Data from calibration experiments. Dim px density plotted vs. cell density determined from cell counts. Bounded exponential fits of individual experiments shown in colored lines. A fit with pooled data ($n = 56$) is shown in black ($[\mathrm{Dim\; Px\; Density}] = 1 - \exp\{-0.00109 \times [\mathrm{Cell\; Density}] - 0.0784\}$, Residual sum-of-squares = 0.1922, Spearman's $\rho$ = 0.949, p $<$ 0.001). (\textbf{b}) Distribution of dim px densities in included radial bands of control dishes ($n = 36 \times 23$). The mean value for the distribution is shown with corresponding predicted cell density using the pooled fit.}
    \label{fig: calib}
    \end{center}

\end{figure}

GFP images from calibration experiments were processed as described in Fig. \ref{fig: imaging}. Data from these calibration experiments is expressed as total (whole dish) dim px density (px/px) vs. cell count per area (cells/sq. mm) in Fig. \ref{fig: calib}a. Grouped data corresponded well to a bounded exponential fit, allowing for the robust conversion of px densities to approximate cell densities. The bounded exponential form of the fit may arise due to changes in the size and morphology of thyroid cells as they approach confluence. To demonstrate that our `calibration curve' was valid for the range of data observed, a probability density distribution of dim px densities in all included radial bands of all P2, P4, and P5 control dishes is shown alongside the fit (Fig. \ref{fig: calib}b). Observed dim px densities fell within the range of the calibration curve. The mean value of this distribution is also indicated. The mean dim px density is 0.374 and the corresponding cell density from the fit is 358 cells/sq. mm. 

\subsection*{Application of 200 kHz induced EMF for 72 h causes moderate EF amplitude-dependent reduction in human thyroid cell density, especially at EF amplitudes exceeding 4 V/cm}

\begin{figure}[hbt!]
    \begin{center}
    \includegraphics{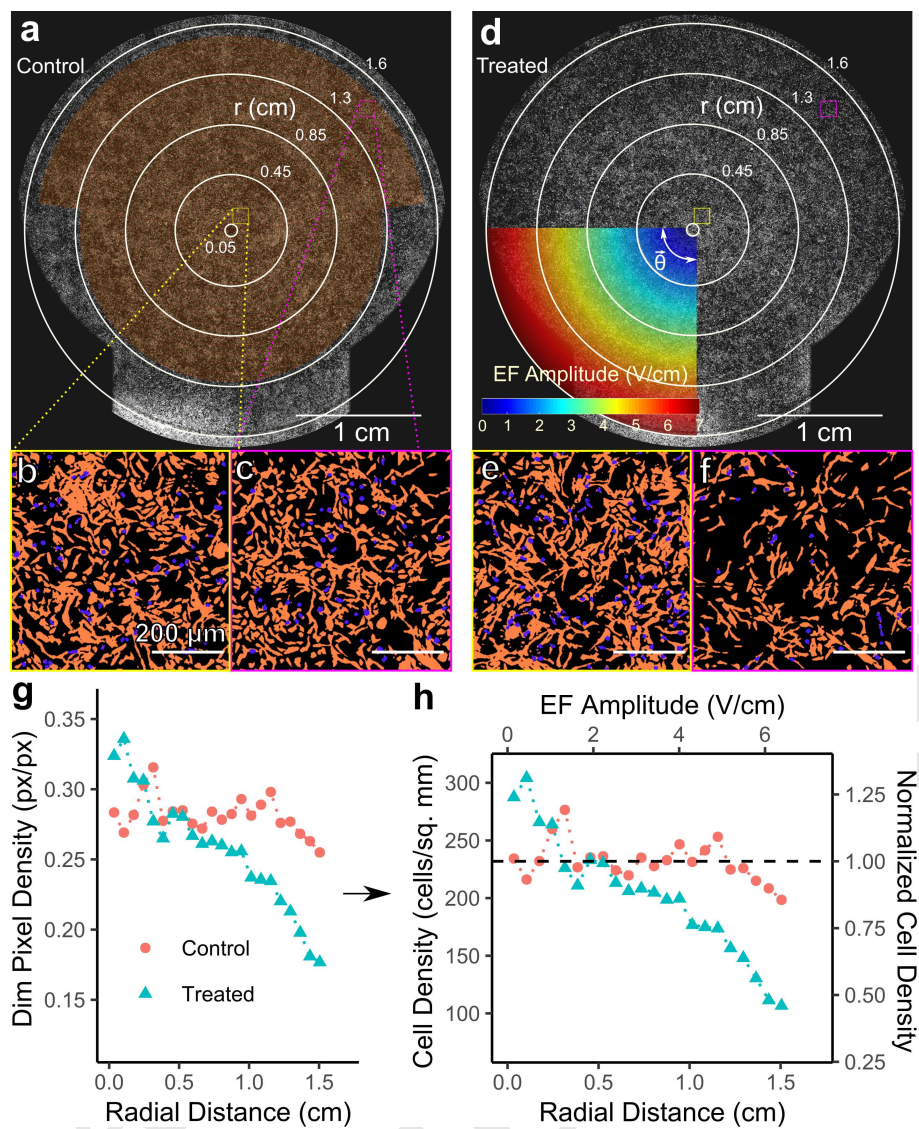}
    \caption{Example of treated vs. control dish and data processing and representation. (\textbf{a} -- \textbf{c}) Example of (a) control dish and zoomed, processed regions near the center (b, yellow) and near the periphery (c, magenta). Whole dish image is adjusted as in Fig. \ref{fig: imaging}i for visualization. (\textbf{d} -- \textbf{f}) Corresponding treated dish with the same zoomed regions. An overlay of the applied 200 kHz EMF amplitude and azimuthal direction is shown. (\textbf{g} -- \textbf{h}) Radially binned data comparing control and treated dishes shown in parts (a - f). (g) Raw dim px density converted to cell density (h, left axis) via the fit in Fig. \ref{fig: calib}, and subsequently normalized to the mean of the control curve (h, right axis).}
    \label{fig: results_eg}
    \end{center}

\end{figure}

A pair of control and treated culture dishes is shown in Figs. \ref{fig: results_eg}a -- b. The displayed images provide an example of a radially dependent cell density reduction in the treated dish. Both a central and peripheral region of the control and treated dishes are shown at 100\% resolution (Figs. \ref{fig: results_eg}b -- c, e -- f) after the described image processing pipeline (Figs. \ref{fig: imaging}a -- d) to demonstrate this radially dependent reduction in dim px density and thus cell density. Conversion from radially binned dim px density to a normalized cell density is also shown in Figs. \ref{fig: results_eg}g -- h using the fit in Fig. \ref{fig: calib}.

\begin{figure}[hbt!]
    \begin{center}
    \includegraphics{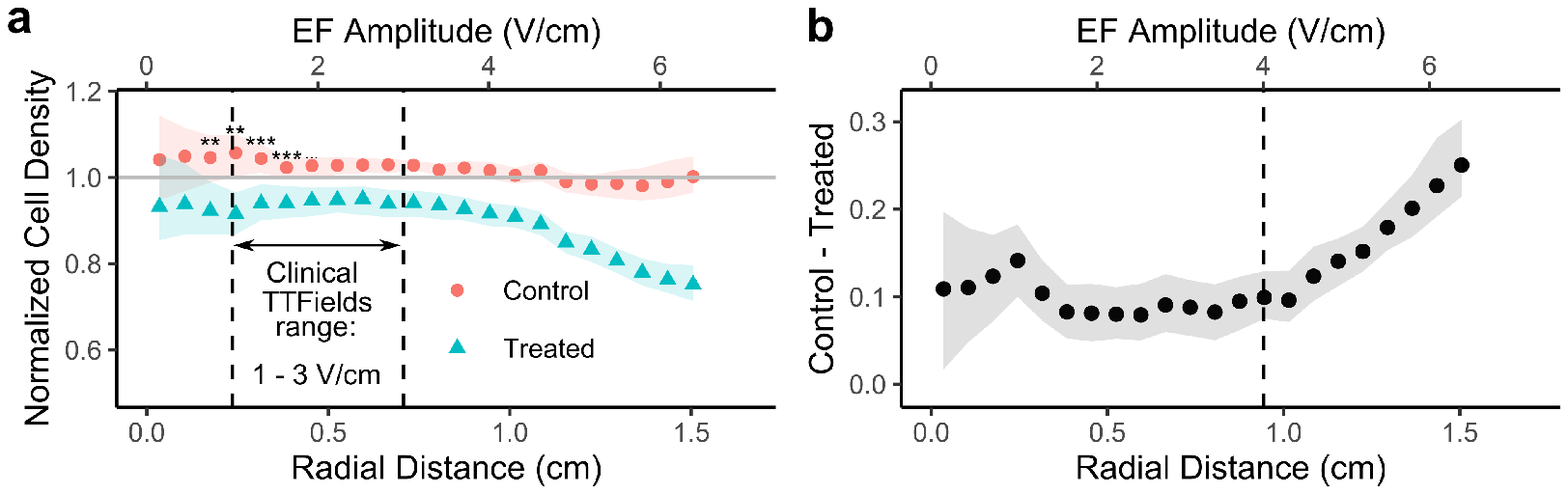}
    \caption{Summary of treatment results presented as normalized cell density curves converted from dim px densities. (\textbf{a}) Control vs. treated curve ($N = 36$). Ribbons = 95\% CI determined by percentile bootstrapping method with 1000 re-samplings. Point-wise t-tests: $** = $ p $< 0.01$, $*** =$ p $< 0.001$. Clinically relevant ``TTFields'' range shown for comparison. (\textbf{b}) Difference between point-wise control vs. treated normalized cell densities bootstrapped in the same way. Dashed line at 4 V/cm indicates apparent regime change.}
    \label{fig: results}
    \end{center}

\end{figure}

Data from all replicates ($N = 36$) was converted to normalized control vs. treated cell density curves as shown in Fig. \ref{fig: results_eg}h. We address these normalized dim px density results first, and bright px density and EthD-1 and PI staining later. Average curves with bootstrapped 95\% confidence intervals (CI) are shown in Fig. \ref{fig: results}a. The bootstrapped difference between curves is shown in Fig. \ref{fig: results}b. Results demonstrate that significant differences in cell density are present throughout the dish, even at EF amplitudes thought to be sub-therapeutic ($<$ 1 V/cm) \cite{Kirson2007}. These differences near the center of the dish cannot be clearly attributed to EF stimulation and could potentially be the result of conditions present throughout the dish, such as the final osmolarity difference between control and treated conditions discussed in Fig. \ref{fig: osmo}, mild temperature gradients (see Fig. \ref{fig: T measurement}, Supp. Sec. 2), or the alternating magnetic field (see Discussion). At higher EF amplitudes ($>$ 4 V/cm) an EF amplitude-dependent effect emerges and the difference between the control and treated conditions increases linearly with EF amplitude from about 10\% to 25\% from 4 to 6.5 V/cm. At lower EF amplitudes in the same dish, a constant difference is instead observed, suggesting the presence of heterogeneous effects in different amplitude regimes. Note, also, that this is a magnitude cell density decrease in the treated condition, suggesting that the peripheral decrease is not a result of cell motility. A motility effect would instead cause a relative increase in central density. Overall, these results suggest a modest effect on human thyroid cell line density by 200 kHz EMF, and, consistent with the literature, the effect is more pronounced at higher EF amplitudes.

\subsection*{Reduction in human thyroid cell density is not sufficiently explained by cell death}

\begin{figure}[hbt!]
    \begin{center}
    \includegraphics{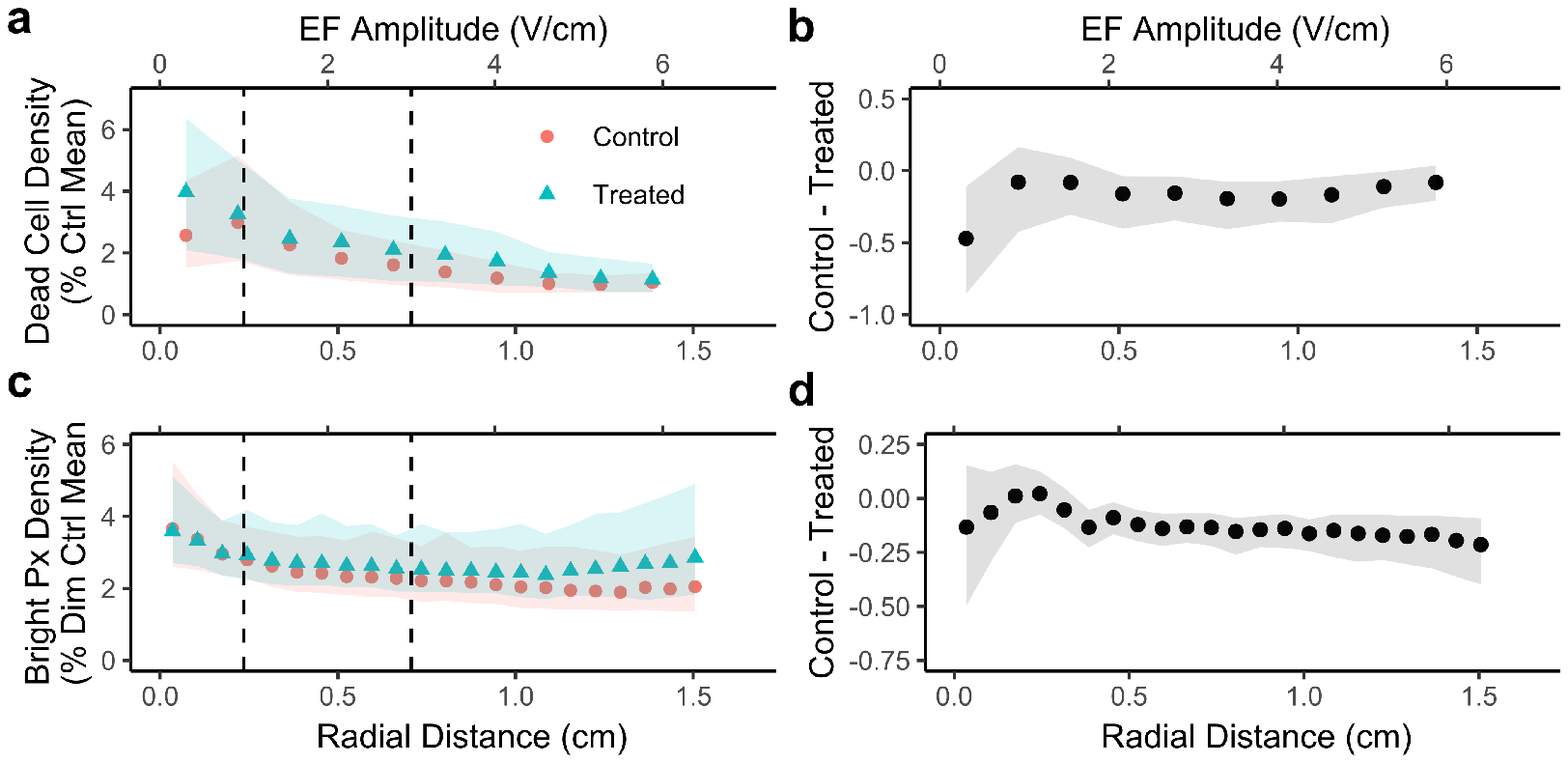}
    \caption{Cell death and bright px density in response to treatment. (\textbf{a} - \textbf{b}) Dead cell density curves and differences constructed from pooled EthD-1 staining ($n = 6$) and PI staining ($n = 6$) results processed as described and expressed as a percent of control mean cell density (358 cells/sq. mm). Ribbons = 95\% CI. No point-wise two-sample independent $t$-tests were significant. (\textbf{c} - \textbf{d}) Bright cell density curves and difference ($n = 36$), expressed as percent of control mean dim px density (0.374 px/px). Again, no point-wise two-sample independent $t$-tests were significant.}
    \label{fig: results_secondary}
    \end{center}

\end{figure}

The observed reduction in cell density can be explained by a reduction in proliferation, an increase in cell death, or some mixture of both mechanisms. To explore their relative roles, bright px density and dead cell density were both quantified, shown in Fig. \ref{fig: results_secondary}. Bright px density, which may correspond to cellular debris, pre-apoptotic cells, mitotic cells, or a rounded cell morphology, was quantified in the same way as dim px density (Fig. \ref{fig: imaging}d) but considering the brighter (violet) segmented population. Dead cell density was quantified as described in the methods. Data from experiments with EthD-1 and PI staining are pooled. Both bright px density (Fig. \ref{fig: results_secondary}c -- d) and dead cell density (Fig. \ref{fig: results_secondary}a -- b) are plotted as a percentage of the control mean values indicated in Fig. \ref{fig: calib} to contextualize the relative contribution of bright px and cell death to the effect observed in Fig. \ref{fig: results}. Results indicate that while the treated condition exhibits some qualitative differences in the radial curves for bright px and dead cell density, these differences are not statistically significant at the replicate number used. Moreover, as a percentage of the control means in Fig. \ref{fig: calib}, the differences are small ($<$ 1\%) compared to the 10 -- 25\% differences observed in Figs. \ref{fig: results}a and b. Bright px and dead cells therefore cannot sufficiently explain the observed effects, especially at higher EF amplitudes. The observed effect appears to be anti-proliferative rather than destructive or morphological based on these preliminary mechanistic explorations.

\section*{Discussion}

In this work, we detail the development of methods to deliver moderate amplitude, intermediate frequency EFs using induced EMFs. We were able to experimentally validate the delivery of a spatially varying EF profile and quantify substantial experimental improvements in thermal and osmolar regulation. We also report novel results of 72 h application of continuous, 200 kHz, 0 to 6.5 V/cm EMF stimulation on a non-cancerous, proliferating human cell line. Results were analyzed using a newly developed image processing pipeline designed specifically to characterize the \textit{in situ} images necessitated by the application of a spatially varying EF ``dose'' profile. Our results suggest that (1) 200 kHz EMF have a moderate maximal effect (< 25\%) on non-cancerous cells in this amplitude range, (2) the effect of 200 kHz EMF on cell density is likely anti-proliferative (i.e., anti-mitotic) rather than destructive, and (3) the effect on cell density increases with increasing EF amplitude in a higher EF regime (> 4 V/cm).

Our exploratory results corroborate and extend prior findings on the effects of ``TTFields'' on proliferating, non-cancerous cell lines. Kirson et al. \cite{Kirson2004} reported that baby hamster kidney (BHK) fibroblasts grown in 0.1\% fetal calf serum exhibited no decrease in growth rate during 24 h application of 1.2 V/cm pk, 100 kHz EFs. Similarly, Jo et al. \cite{Jo2018} reported little to no reduction in cell count for normal human skin cells (CCD-986sk) after 72 h application of up to 1.5 V/cm pk, 200 kHz EFs. Our results agree with these findings. Within and below the therapeutic EF amplitude range of 1 -- 3 V/cm pk-pk, we found that human thyroid cells (Nthy-ori 3-1 Sigma) did not exhibit EF amplitude-dependent growth reduction. We then extend prior findings by showing that, for this cell line, an EF amplitude-dependent effect is observed at $>$ 4 V/cm pk-pk -- which is well beyond the amplitudes used by Kirson et al. \cite{Kirson2004} and Jo et al. \cite{Jo2018}. These results suggest that there may be a range of EF amplitudes within which some non-cancerous cells are unaffected. Further studies may clarify whether such an EF amplitude threshold exists for other non-cancerous cell lines such as CCD-986sk. 

Our findings also provide preliminary support for anti-mitotic mechanisms of action such as the mitotic spindle disruption reported by Giladi et al. \cite{Giladi2015}, rather than cell death-related mechanisms (i.e., autophagy or necroptosis). However, these findings do not necessarily generalize to cancerous cell lines. Indeed, it is known that the effects of alternating EFs are highly variable between cell lines \cite{Kirson2004, Jo2018, Gera2015, Giladi2015, Silginer2017}. Future studies using cancerous cell lines are necessary to substantiate any mechanistic claims. Nonetheless, these results constitute a proof-of-concept for our device and lay the methodological and procedural foundation for future studies of glioma and other cancer cell lines. Importantly, temperature results in Fig. \ref{fig: T measurement} and Fig. S3 confirm that our device can potentially operate at even greater power (and thus EF amplitudes) by simply lowering the coil temperature, providing the capability to investigate the effects of previously unexplored EF amplitudes.

More specifically, our device is well-suited to begin clarifying biophysical mechanisms that can be assessed at an ``endpoint''. Some of the proposed mechanisms can only be studied with simultaneous EF delivery and imaging (e.g., Ref. \cite{Giladi2015}), which is not feasible with the current configuration of our device. Other mechanisms, however, are easily assessed at the experimental endpoint. For example, Kim et al. \cite{Kim2019} reported an autophagy effect in glioblastoma cells mediated by deactivation of the Akt2/miR29b axis. Our device can readily study radially-dependent cell death, as reported via imaging observables here. Complementary genomic or transcriptomic studies can also be performed by merely lifting cells from different radial positions within the dish. Thus, we can begin to validate findings such as those reported by Kim et al. \cite{Kim2019} using our alternative EF delivery platform. As a second example, reported DNA damage effects \cite{Giladi2017, Jo2018} can likewise be assessed by performing the appropriate assays (e.g., an alkaline comet assay) on cells lifted from different radial positions. We reiterate that our device has the unique advantage of each dish representing an internally controlled EF dose titration experiment.

Although treated interchangeably with EFs throughout, it should be noted that using induced EMFs to generate 200 kHz EFs results in an accompanying magnetic field that was not present when using previous delivery methods. At the power levels used, a vertically alternating uniform magnetic field of about 0.6 millitesla is expected to be produced within the coil according to Eq. \eqref{eqn: B}. This magnetic field may cause its own biological effects. Recent research suggests that high gradient and amplitude magnetic fields can affect cell division and resting membrane potentials \cite{Zablotskii2016}. The magnetic field produced here, however, is orders of magnitude smaller than those shown to have biological effects. Moreover, because biological cells and tissues have a diamagnetic susceptibility nearly identical to that of water \cite{Schenck1996} and the magnetic field within a solenoid coil is known to be uniform (and not spatially varying), we do not anticipate that the applied magnetic field will result in any significant magnetic field gradients at interfaces. We acknowledge, however, that magnetic field effects are one possible explanation for the radially uniform level ``offset'' observed in Fig. \ref{fig: results} which cannot be ruled out at this time.

Finally, while the proposed methods offer clear improvements over previous EF delivery methods, our experimental test system is not without practical limitations. The throughput of the system is limited by the use of a single-coil solenoid as the EF applicator. Higher throughput experiments would be bottle-necked by the space requirements of the current platform, which involves a large chiller and induction heater. Another practical drawback is the power requirement; the present 3.19 kW power requirement approaches the battery capacities of industry-leading electric vehicles over a single day ($\approx 75$ kWh). Future developments should attempt to miniaturize the experimental platform and introduce parallelism to enable combinatorial experiments. Alternative coil architectures, chiller loops, multi-coil array designs, controllable capacitor banks, etc., are all worthwhile engineering improvements that could enhance the utility, applicability, and efficiency of this platform. These designs could be motivated by more detailed finite element modeling (FEM) of the various components of this test system, which is beyond the scope of this present work.

\section*{Conclusion}

We report methods to deliver EMFs to cell and tissue cultures using electromagnetic induction, which overcomes prior limitations associated with the delivery of 200 kHz EFs using contacting electrodes. The proposed strategy for inductive EMF delivery eliminates conductive heating of culture dishes and improves temperature control and regulation. Additional design enhancements of our culture dishes improved osmoregulation by mitigating media loss. We report results of stimulation of human thyroid cell line cultures with 200 kHz EMFs exceeding 3 V/cm pk-pk and as high as 6.5 V/cm. Results support a moderate anti-proliferative effect, especially at EF amplitudes exceeding 4 V/cm. The effect is unlikely to be due to cell death based on secondary PI and EthD-1 staining. In summary, the 200 kHz EMF delivery method reported here is a carefully vetted and well-characterized \textit{in vitro} experimental platform for future exploration of the effect of intermediate frequency EMFs on living (and non-living) systems.

\section*{Author Contributions}

R.R., P.J.B. conceived the study. R.R., P.J.B., T.X.C. designed the study. R.H.P., M.G.C., T.P. contributed to electric field delivery and monitoring setup. R.R., H.W., A.G. contributed to cell culture methods. R.R. performed all experiments and acquired data. T.X.C. designed the analysis, analyzed data, and wrote the supplementary material. R.Z.F. contributed to image data analysis. T.X.C., R.R., P.J.B. wrote the manuscript. N.H.W. gave technical advice. M.R.G. and Z.Z. provided cancer and cell biology expertise and contributed to the discussion. P.J.B. supervised the project. All authors edited the manuscript.

\section*{Acknowledgments}

Light microscopy was performed at the NICHD Microscopy \& Imaging Core with the assistance of Dr. Vincent Schram and Mrs. Lynne                 Holtzclaw. We would also like to thank Dr. Nicole Morgan for helpful discussions. The authors declare no conflicts of interest. Support for this study comes from the Intramural Research Program (IRP) of the \textit{Eunice Kennedy Shriver} National Institute of Child Health and Human Development.

\section*{Supporting Material}

Supporting material accompanies this article and contains supplemental sections concerning: \textit{(1)} estimated media loss calculations, \textit{(2)} heat transfer and temperature profile calculations, \textit{(3)} image processing and threshold selection implementation. References \cite{Welty2015, Herning1936, Schwertz1951, Teske2005, Arnold1994, Chen2009, Chapman1990, Wilke1950, Harper2000, Churchill1975, McAdams1985, Bird2006, CRC, Mittal2007} appear in the supporting material. The supporting material can be found by visiting BJ Online at \url{http://www.biophysj.org}.

\bibliography{main}

\begin{thebibliography}{51}
\providecommand{\url}[1]{\texttt{#1}}
\providecommand{\urlprefix}{ }

\bibitem[Cifra et~al.(2011)Cifra, Fields, and Farhadi]{Cifra2011}
Cifra, M., J.~Z. Fields, and A.~Farhadi, 2011.
\newblock Electromagnetic cellular interactions.
\newblock \emph{Progress in Biophysics and Molecular Biology} 105:223 -- 246.
\newblock Muscle Excitation-Contraction Coupling: Elements and Integration.

\bibitem[Kirson et~al.(2004)Kirson, Gurvich, Schneiderman, Dekel, Itzhaki,
  Wasserman, Schatzberger, and Palti]{Kirson2004}
Kirson, E.~D., Z.~Gurvich, R.~Schneiderman, E.~Dekel, A.~Itzhaki, Y.~Wasserman,
  R.~Schatzberger, and Y.~Palti, 2004.
\newblock Disruption of Cancer Cell Replication by Alternating Electric Fields.
\newblock \emph{Cancer Research} 64:3288--3295.

\bibitem[Kirson et~al.(2007)Kirson, Dbal{\'y}, Tovarys, Vymazal, Soustiel,
  Itzhaki, Mordechovich, Steinberg-Shapira, Gurvich, Schneiderman, Wasserman,
  Salzberg, Ryffel, Goldsher, Dekel, and Palti]{Kirson2007}
Kirson, E.~D., V.~Dbal{\'y}, F.~Tovarys, J.~Vymazal, J.~F. Soustiel,
  A.~Itzhaki, D.~Mordechovich, S.~Steinberg-Shapira, Z.~Gurvich,
  R.~Schneiderman, Y.~Wasserman, M.~Salzberg, B.~Ryffel, D.~Goldsher, E.~Dekel,
  and Y.~Palti, 2007.
\newblock Alternating electric fields arrest cell proliferation in animal tumor
  models and human brain tumors.
\newblock \emph{Proceedings of the National Academy of Sciences of the United
  States of America : PNAS} 104:10152--10157.

\bibitem[Batash et~al.(2017)Batash, Asna, Schaffer, Francis, and
  Schaffer]{Batash2017}
Batash, R., N.~Asna, P.~Schaffer, N.~Francis, and M.~Schaffer, 2017.
\newblock Glioblastoma Multiforme, Diagnosis and Treatment; Recent Literature
  Review.
\newblock \emph{Current Medicinal Chemistry} 24:3002--3009.

\bibitem[Gilbert et~al.(2014)Gilbert, Dignam, Armstrong, Wefel, Blumenthal,
  Vogelbaum, Colman, Chakravarti, Pugh, Won, Jeraj, Brown, Jaeckle, Schiff,
  Stieber, Brachman, Werner-Wasik, Tremont-Lukats, Sulman, Aldape, Curran, and
  Mehta]{Gilbert2014}
Gilbert, M.~R., J.~J. Dignam, T.~S. Armstrong, J.~S. Wefel, D.~T. Blumenthal,
  M.~A. Vogelbaum, H.~Colman, A.~Chakravarti, S.~Pugh, M.~Won, R.~Jeraj, P.~D.
  Brown, K.~A. Jaeckle, D.~Schiff, V.~W. Stieber, D.~G. Brachman,
  M.~Werner-Wasik, I.~W. Tremont-Lukats, E.~P. Sulman, K.~D. Aldape, W.~J.~J.
  Curran, and M.~P. Mehta, 2014.
\newblock A randomized trial of bevacizumab for newly diagnosed glioblastoma.
\newblock \emph{New England Journal of Medicine: NEJM} 370:699--708.

\bibitem[Stupp et~al.(2005)Stupp, Mason, van~den Bent, Weller, Fisher,
  Taphoorn, Belanger, Brandes, Marosi, Bogdahn, Curschmann, Janzer, Ludwin,
  Gorlia, Allgeier, Lacombe, Cairncross, Eisenhauer, and Mirimanoff]{Stupp2005}
Stupp, R., W.~P. Mason, M.~J. van~den Bent, M.~Weller, B.~Fisher, M.~J.
  Taphoorn, K.~Belanger, A.~A. Brandes, C.~Marosi, U.~Bogdahn, J.~Curschmann,
  R.~C. Janzer, S.~K. Ludwin, T.~Gorlia, A.~Allgeier, D.~Lacombe, J.~G.
  Cairncross, E.~Eisenhauer, and R.~O. Mirimanoff, 2005.
\newblock Radiotherapy plus Concomitant and Adjuvant Temozolomide for
  Glioblastoma.
\newblock \emph{New England Journal of Medicine: NEJM} 352:987--996.

\bibitem[Stupp et~al.(2012)Stupp, Wong, Kanner, Steinberg, Engelhard, Heidecke,
  Kirson, Taillibert, Liebermann, Dbalý, Ram, Villano, Rainov, Weinberg,
  Schiff, Kunschner, Raizer, Honnorat, Sloan, Malkin, Landolfi, Payer, Mehdorn,
  Weil, Pannullo, Westphal, Smrcka, Chin, Kostron, Hofer, Bruce, Cosgrove,
  Paleologous, Palti, and Gutin]{Stupp2012}
Stupp, R., E.~T. Wong, A.~A. Kanner, D.~Steinberg, H.~Engelhard, V.~Heidecke,
  E.~D. Kirson, S.~Taillibert, F.~Liebermann, V.~Dbalý, Z.~Ram, J.~L. Villano,
  N.~Rainov, U.~Weinberg, D.~Schiff, L.~Kunschner, J.~Raizer, J.~Honnorat,
  A.~Sloan, M.~Malkin, J.~C. Landolfi, F.~Payer, M.~Mehdorn, R.~J. Weil, S.~C.
  Pannullo, M.~Westphal, M.~Smrcka, L.~Chin, H.~Kostron, S.~Hofer, J.~Bruce,
  R.~Cosgrove, N.~Paleologous, Y.~Palti, and P.~H. Gutin, 2012.
\newblock NovoTTF-100A versus physician’s choice chemotherapy in recurrent
  glioblastoma: A randomised phase III trial of a novel treatment modality.
\newblock \emph{European Journal of Cancer} 48:2192 -- 2202.

\bibitem[Stupp et~al.(2017)Stupp, Taillibert, Kanner, Read, Steinberg,
  Lhermitte, Toms, Idbaih, Ahluwalia, Fink, Di~Meco, Lieberman, Zhu,
  Stragliotto, Tran, Brem, Hottinger, Kirson, Lavy-Shahaf, Weinberg, Kim, Paek,
  Nicholas, Bruna, Hirte, Weller, Palti, Hegi, and Ram]{Stupp2017}
Stupp, R., S.~Taillibert, A.~Kanner, W.~Read, D.~Steinberg, B.~Lhermitte,
  S.~Toms, A.~Idbaih, M.~S. Ahluwalia, K.~Fink, F.~Di~Meco, F.~Lieberman, J.-J.
  Zhu, G.~Stragliotto, D.~Tran, S.~Brem, A.~Hottinger, E.~D. Kirson,
  G.~Lavy-Shahaf, U.~Weinberg, C.-Y. Kim, S.-H. Paek, G.~Nicholas, J.~Bruna,
  H.~Hirte, M.~Weller, Y.~Palti, M.~E. Hegi, and Z.~Ram, 2017.
\newblock Effect of Tumor-Treating Fields Plus Maintenance Temozolomide vs
  Maintenance Temozolomide Alone on Survival in Patients With Glioblastoma: A
  Randomized Clinical Trial.
\newblock \emph{JAMA} 318:2306--2316.

\bibitem[Ceresoli et~al.(2019)Ceresoli, Aerts, Dziadziuszko, Ramlau, Cedres,
  van Meerbeeck, Mencoboni, Planchard, Chella, Crin{\`o}, Krzakowski,
  R{\"u}ssel, Maconi, Gianoncelli, and Grosso]{Ceresoli2019}
Ceresoli, G.~L., J.~G. Aerts, R.~Dziadziuszko, R.~Ramlau, S.~Cedres, J.~P. van
  Meerbeeck, M.~Mencoboni, D.~Planchard, A.~Chella, L.~Crin{\`o},
  M.~Krzakowski, J.~R{\"u}ssel, A.~Maconi, L.~Gianoncelli, and F.~Grosso, 2019.
\newblock Tumour Treating Fields in combination with pemetrexed and cisplatin
  or carboplatin as first-line treatment for unresectable malignant pleural
  mesothelioma (STELLAR): a multicentre, single-arm phase 2 trial.
\newblock \emph{The Lancet Oncology} 20:1702--1709.

\bibitem[Gera et~al.(2015)Gera, Yang, Holtzman, Lee, Wong, and
  Swanson]{Gera2015}
Gera, N., A.~Yang, T.~S. Holtzman, S.~X. Lee, E.~T. Wong, and K.~D. Swanson,
  2015.
\newblock Tumor Treating Fields Perturb the Localization of Septins and Cause
  Aberrant Mitotic Exit.
\newblock \emph{PLOS ONE} 10:1--20.

\bibitem[Tuszynski et~al.(2016)Tuszynski, Wenger, Friesen, and
  Preto]{Tuszynski2016}
Tuszynski, J.~A., C.~Wenger, D.~E. Friesen, and J.~Preto, 2016.
\newblock An Overview of Sub-Cellular Mechanisms Involved in the Action of
  TTFields.
\newblock \emph{International Journal of Environmental Research and Public
  Health} 13:1128.

\bibitem[{Wenger} et~al.(2015){Wenger}, {Giladi}, {Bomzon}, {Salvador},
  {Basser}, and {Miranda}]{Wenger2015a}
{Wenger}, C., M.~{Giladi}, Z.~{Bomzon}, R.~{Salvador}, P.~J. {Basser}, and
  P.~C. {Miranda}, 2015.
\newblock Modeling Tumor Treating Fields (TTFields) application in single cells
  during metaphase and telophase.
\newblock \emph{In} 2015 37th Annual International Conference of the IEEE
  Engineering in Medicine and Biology Society (EMBC). 6892--6895.

\bibitem[Wenger et~al.(2015)Wenger, Salvador, Basser, and Miranda]{Wenger2015b}
Wenger, C., R.~Salvador, P.~J. Basser, and P.~C. Miranda, 2015.
\newblock The electric field distribution in the brain during TTFields therapy
  and its dependence on tissue dielectric properties and anatomy: a
  computational study.
\newblock \emph{Physics in medicine and biology} 60:7339--7357.

\bibitem[{Wenger} et~al.(2018){Wenger}, {Miranda}, {Salvador}, {Thielscher},
  {Bomzon}, {Giladi}, {Mrugala}, and {Korshoej}]{Wenger2018}
{Wenger}, C., P.~C. {Miranda}, R.~{Salvador}, A.~{Thielscher}, Z.~{Bomzon},
  M.~{Giladi}, M.~M. {Mrugala}, and A.~R. {Korshoej}, 2018.
\newblock A Review on Tumor-Treating Fields (TTFields): Clinical Implications
  Inferred From Computational Modeling.
\newblock \emph{IEEE Reviews in Biomedical Engineering} 11:195--207.

\bibitem[Davies et~al.(2013)Davies, Weinberg, and Palti]{Davies2013}
Davies, A.~M., U.~Weinberg, and Y.~Palti, 2013.
\newblock Tumor treating fields: a new frontier in cancer therapy.
\newblock \emph{Annals of the New York Academy of Sciences} 1291:86--95.

\bibitem[Bomzon et~al.(2016)Bomzon, Wenger, Giladi, Urman, Schneiderman,
  Voloshin~Sela, Porat, Munster, Blat, sherbo, Weinberg, Kirson, Miranda,
  Wasserman, and Palti]{Bomzon2016}
Bomzon, Z., C.~Wenger, M.~Giladi, N.~Urman, R.~S. Schneiderman,
  T.~Voloshin~Sela, Y.~Porat, M.~Munster, R.~Blat, s.~sherbo, U.~Weinberg,
  E.~Kirson, P.~C. Miranda, Y.~Wasserman, and Y.~Palti, 2016.
\newblock Quantifying the Effect of Electric Fields in the Frequency Range of
  100-500 khz on Mitotic Spindle Structures.
\newblock \emph{Biophysical Journal} 110:619a.

\bibitem[Berkelmann et~al.(2019)Berkelmann, Bader, Meshksar, Dierks,
  Hatipoglu~Majernik, Krauss, Schwabe, Manteuffel, and
  Ngezahayo]{Berkelmann2019}
Berkelmann, L., A.~Bader, S.~Meshksar, A.~Dierks, G.~Hatipoglu~Majernik, J.~K.
  Krauss, K.~Schwabe, D.~Manteuffel, and A.~Ngezahayo, 2019.
\newblock Tumour-treating fields (TTFields): Investigations on the mechanism of
  action by electromagnetic exposure of cells in telophase/cytokinesis.
\newblock \emph{Scientific Reports} 9:7362.

\bibitem[{Li} et~al.(2020){Li}, {Yang}, and {Rubinsky}]{Li2020}
{Li}, X., F.~{Yang}, and B.~{Rubinsky}, 2020.
\newblock A Theoretical Study on the Biophysical Mechanisms by Which Tumor
  Treating Fields Affect Tumor Cells during Mitosis.
\newblock \emph{IEEE Transactions on Biomedical Engineering} 1--1.

\bibitem[Kessler et~al.(2018)Kessler, Frömbling, Gross, Hahn, Dzokou,
  Ernestus, Löhr, and Hagemann]{Kessler2018}
Kessler, A.~F., G.~E. Frömbling, F.~Gross, M.~Hahn, W.~Dzokou, R.-I. Ernestus,
  M.~Löhr, and C.~Hagemann, 2018.
\newblock Effects of tumor treating fields ({TTFields}) on glioblastoma cells
  are augmented by mitotic checkpoint inhibition.
\newblock \emph{Cell Death Discov} 4:12--12.

\bibitem[Santelices et~al.(2017)Santelices, Friesen, Bell, Hough, Xiao, Kalra,
  Kar, Freedman, Rezania, Lewis, Shankar, and Tuszynski]{Santelices2017}
Santelices, I.~B., D.~E. Friesen, C.~Bell, C.~M. Hough, J.~Xiao, A.~Kalra,
  P.~Kar, H.~Freedman, V.~Rezania, J.~D. Lewis, K.~Shankar, and J.~A.
  Tuszynski, 2017.
\newblock Response to Alternating Electric Fields of Tubulin Dimers and
  Microtubule Ensembles in Electrolytic Solutions.
\newblock \emph{Scientific Reports} 7:9594.

\bibitem[Chang et~al.(2018)Chang, Patel, Pohling, Young, Song, Flores, Zeng,
  Joubert, Arami, Natarajan, Sinclair, and Gambhir]{Chang2018}
Chang, E., C.~B. Patel, C.~Pohling, C.~Young, J.~Song, T.~A. Flores, Y.~Zeng,
  L.-M. Joubert, H.~Arami, A.~Natarajan, R.~Sinclair, and S.~S. Gambhir, 2018.
\newblock Tumor treating fields increases membrane permeability in glioblastoma
  cells.
\newblock \emph{Cell Death Discovery} 4:113.

\bibitem[Giladi et~al.(2017)Giladi, Munster, Schneiderman, Voloshin, Porat,
  Blat, Zielinska-Chomej, Hååg, Bomzon, Kirson, Weinberg, Viktorsson,
  Lewensohn, and Palti]{Giladi2017}
Giladi, M., M.~Munster, R.~S. Schneiderman, T.~Voloshin, Y.~Porat, R.~Blat,
  K.~Zielinska-Chomej, P.~Hååg, Z.~Bomzon, E.~D. Kirson, U.~Weinberg,
  K.~Viktorsson, R.~Lewensohn, and Y.~Palti, 2017.
\newblock Tumor treating fields ({TTFields}) delay {DNA} damage repair
  following radiation treatment of glioma cells.
\newblock \emph{Radiat Oncol} 12:206--206.

\bibitem[Jo et~al.(2018)Jo, Hwang, Jin, Sung, Jeong, Baek, Cho, Kim, and
  Yoon]{Jo2018}
Jo, Y., S.-G. Hwang, Y.~B. Jin, J.~Sung, Y.~K. Jeong, J.~H. Baek, J.-M. Cho,
  E.~H. Kim, and M.~Yoon, 2018.
\newblock Selective toxicity of tumor treating fields to melanoma: an in vitro
  and in vivo study.
\newblock \emph{Cell Death Discovery} 4:46.

\bibitem[Silginer et~al.(2017)Silginer, Weller, Stupp, and Roth]{Silginer2017}
Silginer, M., M.~Weller, R.~Stupp, and P.~Roth, 2017.
\newblock Biological activity of tumor-treating fields in preclinical glioma
  models.
\newblock \emph{Cell Death Dis} 8:e2753--e2753.

\bibitem[Kim et~al.(2019)Kim, Jo, Sai, Park, Kim, Kim, Lee, Cho, Kwak, Baek,
  Jeong, Song, Yoon, and Hwang]{Kim2019}
Kim, E.~H., Y.~Jo, S.~Sai, M.-J. Park, J.-Y. Kim, J.~S. Kim, Y.-J. Lee, J.-M.
  Cho, S.-Y. Kwak, J.-H. Baek, Y.~K. Jeong, J.-Y. Song, M.~Yoon, and S.-G.
  Hwang, 2019.
\newblock Tumor-treating fields induce autophagy by blocking the Akt2/miR29b
  axis in glioblastoma cells.
\newblock \emph{Oncogene} 38:6630--6646.

\bibitem[Neuhaus et~al.(2019)Neuhaus, Zirjacks, Ganser, Klumpp, Sch{\"u}ler,
  Zips, Eckert, and Huber]{Neuhaus2019}
Neuhaus, E., L.~Zirjacks, K.~Ganser, L.~Klumpp, U.~Sch{\"u}ler, D.~Zips,
  F.~Eckert, and S.~M. Huber, 2019.
\newblock Alternating Electric Fields (TTFields) Activate Ca(v)1.2 Channels in
  Human Glioblastoma Cells.
\newblock \emph{Cancers} 11:110.

\bibitem[Giladi et~al.(2015)Giladi, Schneiderman, Voloshin, Porat, Munster,
  Blat, Sherbo, Bomzon, Urman, Itzhaki, Cahal, Shteingauz, Chaudhry, Kirson,
  Weinberg, and Palti]{Giladi2015}
Giladi, M., R.~S. Schneiderman, T.~Voloshin, Y.~Porat, M.~Munster, R.~Blat,
  S.~Sherbo, Z.~Bomzon, N.~Urman, A.~Itzhaki, S.~Cahal, A.~Shteingauz,
  A.~Chaudhry, E.~D. Kirson, U.~Weinberg, and Y.~Palti, 2015.
\newblock Mitotic Spindle Disruption by Alternating Electric Fields Leads to
  Improper Chromosome Segregation and Mitotic Catastrophe in Cancer Cells.
\newblock \emph{Scientific reports} 5:18046--18046.

\bibitem[Porat et~al.(2017)Porat, Giladi, Schneiderman, Blat, Shteingauz,
  Zeevi, Munster, Voloshin, Kaynan, Tal, Kirson, Weinberg, and
  Palti]{Porat2017}
Porat, Y., M.~Giladi, R.~S. Schneiderman, R.~Blat, A.~Shteingauz, E.~Zeevi,
  M.~Munster, T.~Voloshin, N.~Kaynan, O.~Tal, E.~D. Kirson, U.~Weinberg, and
  Y.~Palti, 2017.
\newblock Determining the {Optimal} {Inhibitory} {Frequency} for {Cancerous}
  {Cells} {Using} {Tumor} {Treating} {Fields} ({TTFields}).
\newblock \emph{JoVE} e55820.

\bibitem[Finan and Guilak(2010)]{Finan2010}
Finan, J.~D., and F.~Guilak, 2010.
\newblock The effects of osmotic stress on the structure and function of the
  cell nucleus.
\newblock \emph{Journal of cellular biochemistry} 109:460--467.

\bibitem[Radmaneshfar et~al.(2013)Radmaneshfar, Kaloriti, Gustin, Gow, Brown,
  Grebogi, Romano, and Thiel]{Radmaneshfar2013}
Radmaneshfar, E., D.~Kaloriti, M.~C. Gustin, N.~A.~R. Gow, A.~J.~P. Brown,
  C.~Grebogi, M.~C. Romano, and M.~Thiel, 2013.
\newblock From {START} to {FINISH}: {The} {Influence} of {Osmotic} {Stress} on
  the {Cell} {Cycle}.
\newblock \emph{PLOS ONE} 8:1--14.

\bibitem[Fragniere et~al.(2019)Fragniere, Stott, Fazal, Andreasen, Scott, and
  Barker]{Fragniere2019}
Fragniere, A. M.~C., S.~R.~W. Stott, S.~V. Fazal, M.~Andreasen, K.~Scott, and
  R.~A. Barker, 2019.
\newblock Hyperosmotic stress induces cell-dependent aggregation of
  a-synuclein.
\newblock \emph{Scientific Reports} 9:2288.

\bibitem[Bickford and Fremming(1956)]{Bickford1965}
Bickford, R., and B.~Fremming, 1956.
\newblock Neuronal stimulation by pulsed magnetic fields in animals and man.
\newblock \emph{In} Digest of the 6th International Conference in Medical
  Electronics and Biological Engineering. IFMBE, 112.

\bibitem[Hassanpour et~al.(2015)Hassanpour, Samadiani, and
  Salehi]{Hassanpour2015}
Hassanpour, H., N.~Samadiani, and S.~M.~M. Salehi, 2015.
\newblock Using morphological transforms to enhance the contrast of medical
  images.
\newblock \emph{The Egyptian Journal of Radiology and Nuclear Medicine} 46:481
  -- 489.

\bibitem[Soille(2004)]{Soille2004}
Soille, P., 2004.
\newblock Morphological Image Analysis : Principles and Applications.
\newblock Springer Berlin Heidelberg, Berlin, Heidelberg.

\bibitem[Otsu(1979)]{Otsu1979}
Otsu, N., 1979.
\newblock A Threshold Selection Method from Gray-Level Histograms.
\newblock \emph{IEEE Transactions on Systems, Man, and Cybernetics} 9:62--66.

\bibitem[Zablotskii et~al.(2016)Zablotskii, Polyakova, Lunov, and
  Dejneka]{Zablotskii2016}
Zablotskii, V., T.~Polyakova, O.~Lunov, and A.~Dejneka, 2016.
\newblock How a High-Gradient Magnetic Field Could Affect Cell Life.
\newblock \emph{Scientific Reports} 6:37407.

\bibitem[Schenck(1996)]{Schenck1996}
Schenck, J.~F., 1996.
\newblock The role of magnetic susceptibility in magnetic resonance imaging:
  MRI magnetic compatibility of the first and second kinds.
\newblock \emph{Medical Physics} 23:815--850.

\bibitem[Welty et~al.(2015)Welty, Wicks, Rorrer, and Wilson]{Welty2015}
Welty, J., C.~E. Wicks, G.~L. Rorrer, and R.~E. Wilson, 2015.
\newblock Fundamentals of momentum, heat, and mass transfer.
\newblock Wiley, Hoboken, NJ.

\bibitem[Herning and Zipperer(1936)]{Herning1936}
Herning, F., and L.~Zipperer, 1936.
\newblock Calculation of the viscosity of technical gas mixtures from the
  viscosity of the individual gases.
\newblock \emph{Gas u. Wasserfach} 79:69.

\bibitem[Schwertz and Brow(1951)]{Schwertz1951}
Schwertz, F.~A., and J.~E. Brow, 1951.
\newblock Diffusivity of {Water} {Vapor} in {Some} {Common} {Gases}.
\newblock \emph{The Journal of Chemical Physics} 19:640--646.

\bibitem[Teske et~al.(2005)Teske, Vogel, and Bich]{Teske2005}
Teske, V., E.~Vogel, and E.~Bich, 2005.
\newblock Viscosity {Measurements} on {Water} {Vapor} and {Their} {Evaluation}.
\newblock \emph{Journal of Chemical \& Engineering Data} 50:2082--2087.

\bibitem[Arnold and Fuhr(1994)]{Arnold1994}
Arnold, W., and G.~Fuhr, 1994.
\newblock Increasing the permittivity and conductivity of cellular
  electromanipulation media.
\newblock \emph{In} Proceedings of 1994 {IEEE} {Industry} {Applications}
  {Society} {Annual} {Meeting}. volume~2, 1470--1476 vol.2.

\bibitem[Chen et~al.(2009)Chen, Jiang, Vernier, Wu, and Gundersen]{Chen2009}
Chen, M.-T., C.~Jiang, P.~T. Vernier, Y.-H. Wu, and M.~A. Gundersen, 2009.
\newblock Two-dimensional nanosecond electric field mapping based on cell
  electropermeabilization.
\newblock \emph{PMC Biophysics} 2:9.

\bibitem[Chapman et~al.(1990)Chapman, Cowling, and Burnett]{Chapman1990}
Chapman, S., T.~G. Cowling, and D.~Burnett, 1990.
\newblock The {Mathematical} {Theory} of {Non}-uniform {Gases}: {An} {Account}
  of the {Kinetic} {Theory} of {Viscosity}, {Thermal} {Conduction} and
  {Diffusion} in {Gases}.
\newblock Cambridge University Press.

\bibitem[Wilke(1950)]{Wilke1950}
Wilke, C.~R., 1950.
\newblock A {Viscosity} {Equation} for {Gas} {Mixtures}.
\newblock \emph{The Journal of Chemical Physics} 18:517--519.

\bibitem[Harper(2000)]{Harper2000}
Harper, C.~A., 2000.
\newblock Modern Plastics Handbook.
\newblock McGraw-Hill Professional.

\bibitem[Churchill and Chu(1975)]{Churchill1975}
Churchill, S.~W., and H.~H.~S. Chu, 1975.
\newblock Correlating equations for laminar and turbulent free convection from
  a horizontal cylinder.
\newblock \emph{International Journal of Heat and Mass Transfer} 18:1049 --
  1053.

\bibitem[McAdams(1985)]{McAdams1985}
McAdams, W.~H., 1985.
\newblock Heat Transmission.
\newblock Krieger Pub Co.

\bibitem[Bird et~al.(2006)Bird, Stewart, and Lightfoot]{Bird2006}
Bird, R.~B., W.~E. Stewart, and E.~N. Lightfoot, 2006.
\newblock Transport Phenomena, Revised 2nd Edition.
\newblock John Wiley \& Sons, Inc.

\bibitem[Haynes(2011)]{CRC}
Haynes, W.~M., 2011.
\newblock CRC Handbook of Chemistry and Physics, 92nd Edition.
\newblock CRC Press.

\bibitem[Mittal et~al.(2007)Mittal, Rosenthal, and Voldman]{Mittal2007}
Mittal, N., A.~Rosenthal, and J.~Voldman, 2007.
\newblock {nDEP} microwells for single-cell patterning in physiological media.
\newblock \emph{Lab on a Chip} 7:1146.

\end{thebibliography}

\end{document}